\documentclass[aps,pra,superscriptaddress,showpacs]{revtex4}
\usepackage{graphicx}
\usepackage{epstopdf,epsfig}
\usepackage{amsmath}
\usepackage{amsfonts}
\usepackage{times}
\begin{document}

\title{Ultracold atoms in $U(2)$ non--Abelian gauge potentials preserving the Landau levels}

\author{Michele Burrello}
\affiliation{SISSA, Via Bonomea 265, I-34136, Trieste, Italy}
\affiliation{INFN, Sezione di Trieste, I-34127 Trieste, Italy}

\author{Andrea Trombettoni}
\affiliation{SISSA, Via Bonomea 265, I-34136, Trieste, Italy}
\affiliation{INFN, Sezione di Trieste, I-34127 Trieste, Italy}

\date{\today}

\begin{abstract}
We study ultracold atoms subjected to $U(2)$ non--Abelian potentials: 
we consider gauge potentials having, in the Abelian limit, 
degenerate Landau levels and 
we then investigate the effect of general homogeneous non--Abelian terms. 
The conditions under which the structure of degenerate 
Landau levels is preserved are classified and discussed. 
The typical gauge potentials preserving the Landau levels are characterized 
by a fictitious magnetic field and by an effective spin--orbit interaction, 
e.g. obtained through the rotation of 
two--dimensional atomic gases coupled with a tripod scheme. 
The single--particle energy spectrum can be exactly determined 
for a class of gauge potentials, whose physical implementation is  
explicitly discussed. The corresponding Landau levels are deformed 
by the non--Abelian contribution of the potential and their spin degeneracy is split. 
The related deformed quantum Hall states for fermions and bosons 
(in the presence of strong intra--species interaction) 
are determined far from and at the degeneracy points of the Landau levels. 
A discussion of the effect of the angular momentum is presented, as well as 
results for $U(3)$ gauge potentials. 
\end{abstract}

\maketitle

\section{Introduction}
\label{introd}

The recent experimental realization of artificial magnetic and electric 
fields acting on neutral atoms \cite{lin09,lin11} opened the way to simulate 
ultracold many--body systems in controllable (static) electro--magnetic fields 
\cite{bloch08}: these synthetic fields can be implemented using 
spatially dependent optical couplings between internal states of the atoms. 
This technique has been applied so far not only to single--component Bose gases 
\cite{lin09,lin11,lin09_2}, but also to Bose--Einstein condensates with 
two components \cite{lin11_2,fu11}: in \cite{lin11_2}, using a suitable 
spatial variation and time dependence of the effective
vector potential, a spin--orbit coupling has been realized. 
Besides, the optically--induced effective gauge potential and spin--orbit 
couplings may be experimentally investigated in the future also 
in ultracold fermionic gases and Bose--Fermi mixtures. 
Such a possibility, eventually together with the use of rotating traps 
\cite{cooper08} and/or the application of 
more general multipod schemes \cite{dalibard10}, 
envisions the concrete opportunity of manipulating and 
studying interacting ultracold systems in a vast class of 
simulated magnetic and electric fields.

An important perspective motivated by the realization 
of synthetic magnetic fields is given by the simulation of quantum Hall 
physics with ultracold atoms. As it is well known, 
the effect of a rotation on a neutral ultracold 
gas is equivalent to that of a magnetic field on a system of charged particles 
\cite{leggett06}: since the first experiments with Bose gases in rotating traps 
\cite{pitaevskii03,pethick08}, this fact called for 
 the possibility to realize Laughlin states 
and other quantum Hall states using two--dimensional 
strongly interacting ultracold gases 
in rapidly rotating traps (see the reviews 
\cite{bloch08,cooper08,viefers08,fetter09}). 
However the achievement of Laughlin states and quantum Hall regimes  
with rotating gases turned out to be 
in general not an easy task, 
since for typical experimental numbers 
the rotation frequency $\Omega$ should be very close to the trap 
frequency $\omega$ \cite{cooper08,fetter09}. 
The recent realization of synthetic magnetic fields 
using spatially dependent optical couplings adds then new technical 
possibilities in view of the experimental realization of quantum Hall states. 
Moreover the first experimental evidences of these strongly--correlated states have been produced 
for a set of a few atoms of $^{87}$Rb loaded 
in time--modulated optical lattices \cite{chu10}.

Another very active line of research is driven by the possibility to apply 
controllable artificial magnetic fields on a two-- (eventually multi--) 
component ultracold 
gas, in particular engineering tunable spin--orbit interactions. 
Spin--orbit coupled Bose--Einstein condensates 
have been experimentally realized 
\cite{lin11_2,fu11}: in \cite{lin11_2} the spin--orbit coupling had equal Rashba and 
Dresselhaus strengths, 
but a much wider class of spin--orbit couplings may be realized 
\cite{ruse05,stanescu07,dalibard10}. 
At low temperatures, a spin--orbit coupled Bose gas 
can condense into degenerate minima at finite momenta 
\cite{stanescu08}; moreover, modifying the interparticle interaction, a spin stripe phase 
may appear \cite{zhai10}. Equally interesting would be the realization 
of spin--orbit coupled ultracold fermionic gases: 
in particular, a Fermi superfluid in the presence 
of a tunable spin--orbit could be studied. The properties of the superconducting state 
in the presence of a spin--orbit interaction lifting the spin degeneracy, 
with mixed singlet and triplet pairings, include anisotropic 
spin magnetic susceptibility and finite Knight shift at zero temperature 
\cite{gorkov01}; such properties have been actively studied in the last decade 
\cite{frigeri04,cappelluti07,ghosh10,fu10} and, very recently, the possible realization 
and study of their counterpart in atomic Fermi superfluids 
attracted significant attention \cite{sato09,kubasiak10,shenoy11,iskin11,gong11,han11,chen11}.

Spin--orbit couplings acting on a two--component ultracold 
gas are a particular example 
of the so--called {\it non--Abelian} gauge potentials: the 
single--particle Hamiltonian has in general a term 
proportional to $(\vec{p}+\vec{A})^2$, where the vector 
$\vec{A}$ has non--commuting components, e.g. $\left[A_x,A_y \right] \neq 0$. 
It is intended that the vector potential is a matrix 
(e.g., $2 \times 2$ for a two--component gas) 
and, in general, it can have a spatial dependence. 
In the tripod scheme, three internal quasi-degenerate states are coupled with a fourth and the
two resulting degenerate dark states are subjected to an effective non--Abelian gauge potential 
\cite{ruse05,juze10}. This scheme can be extended to the tetrapod 
configuration \cite{juze10}; a discussion of more general multipod setups is presented 
in \cite{dalibard10}. $U(2)$ vector potentials acting on ultracold atoms 
in optical lattices can be implemented using 
laser assisted tunneling depending on the hyperfine levels \cite{lewen05}; besides, 
effective non--Abelian gauge 
potentials have been also discussed in cavity QED models \cite{larson09}. 

Several properties of ultracold atoms in artificial 
non--Abelian gauge potentials have been recently studied  
\cite{clark06,santos07,lu07,clark08,juze08,santos08,larson09a,goldman09,lewen09,bermudez10,trombettoni10,schoutens11,palmer11}. If a general non--Abelian term 
is added to an Abelian potential exhibiting (eventually degenerate) 
Landau levels, then the coupling between the internal degrees of freedom 
breaks the degeneracy of the Landau levels: e.g., in \cite{santos08} a matrix 
generalization of the Landau gauge has been considered, and the consequent 
disappearance of the Landau level structure for large non--Abelian terms 
investigated. The possibility of having anomalous 
quantum Hall effects in suitable 
artificial gauge potentials  
has been also addressed \cite{goldman09,lewen09,bermudez10}. 
The Landau level spectrum in a spatially constant non--Abelian vector potential 
in planar and spherical geometries has been discussed in \cite{schoutens11}, 
showing that the adiabatic insertion of a non--Abelian flux 
in a spin--polarized quantum Hall state leads to the 
formation of charged spin--textures. 

An appealing issue regarding two--dimensional ultracold atomic gases subjected 
to controllable non--Abelian gauge potentials concerns the possibility to use 
them to simulate and manipulate ground--states having non--Abelian excitations. 
This interest is due to the highly non--trivial 
topological properties of such correlated 
states \cite{stern08,nayak08} and to their relevance for the 
topological quantum computation schemes \cite{nayak08}. However, even if the 
Landau level structure is not broken, in general the excitations of the system 
remain Abelian \cite{lewen09,trombettoni10,palmer11}. In \cite{palmer11}, using 
exact diagonalization, the fractional quantum Hall effect for two--dimensional 
interacting bosons 
in the presence of a non--Abelian gauge field 
(in addition to the usual Abelian magnetic field) was studied, obtaining 
that for small non--Abelian fields one has 
a single internal state quantum Hall system, whereas for stronger 
fields there is a two internal state behaviour 
(or the complete absence of Hall plateaus). 

In \cite{trombettoni10} the Landau levels and the quantum Hall states were 
studied for a two--component two--dimensional 
gas subjected to the non--Abelian gauge potential 
$A_x=q\sigma_x-\frac{B}{2}y$, $A_y=q\sigma_y+\frac{B}{2}x$ (where the 
$\sigma$'s are the Pauli matrices). The reasons for choosing such potential  
were the following: 
{\it i)} for $q=0$, an artificial magnetic field 
$B$ (perpendicular to the plane $xy$) 
is applied to both the components and the Landau levels are doubly degenerate; 
{\it ii)} for finite $q$, i.e. for a finite non--Abelian term, 
the Landau levels are preserved, 
but their degeneracy is lift; {\it iii)} closed analytical expressions 
for the single--particle energy spectrum can be found 
(see \cite{rashba84}); 
{\it iv)} last but not least, one can choose the parameters of the 
laser pulses in a tripod scheme in such way that it can be 
experimentally implementable. 
For both bosons (in the presence of a strong intra--species interaction) 
and fermions, explicit expressions for the ground--states 
were obtained, showing that deformed Laughlin states  - with Abelian 
excitations - appear; however, ground--states with non--Abelian excitations 
emerge at the points (e.g., $q^2=3B$) in which different Landau levels have the same 
energy. 

The goal of this paper is two--fold. From one side we 
investigate the effect of general homogeneous non--Abelian terms added to the usual 
Abelian magnetic field (assumed equal for both the components, so that 
the Landau levels are degenerate in the Abelian limit), in order 
to discuss and classify under which conditions 
the structure of degenerate Landau levels is preserved. 
From the the other side we present a detailed discussion of the 
corresponding many--particle quantum Hall states derived in \cite{trombettoni10}, 
providing additional results and a discussion of the effect of an 
angular momentum term. We also consider 
$U(3)$ non--Abelian potentials giving rise to lines (in place of points) 
of degeneracy.

The plan of the paper is the following: in Section \ref{intro} 
we study the single--particle Hamiltonian and the Landau levels 
of a two--component two--dimensional 
Bose gas in an artificial $U(2)$ gauge potential, having doubly degenerate 
Landau levels in the Abelian limit and a non-Abelian $SU(2)$ 
gauge potential independent on the position. We argue that 
there are only three classes of (quadratic) Hamiltonians preserving the 
degeneracy of the Landau levels and 
giving rise to an analytically defined Landau level structure, 
where the eigenstates are expressed as a finite linear combination 
of the eigenstates of a particle in a magnetic field. 
The first class corresponds to an Abelian $U(1)\times U(1)$ 
gauge potential and it refers to uncoupled internal states, 
whereas the other two are characterized by a truly non--Abelian 
$U(2)$ gauge potential and correspond to different kinds 
of Jaynes--Cummings models. Then we focus on one of these 
Jaynes--Cummings classes which 
is gauge equivalent both to a Rashba and to a Dresselhaus 
spin--orbit coupling: after discussing in Section \ref{sectripod} how this 
non--Abelian gauge potential can be implemented in a rotating tripod scheme, 
we analyze the Landau level spectrum in Section \ref{oneparticle}, 
where we also present results for $U(3)$ gauge potentials 
acting on a three--component gas. In Section \ref{twobody} we write 
the deformed Laughlin states in the presence of two--body interactions, while in 
Section \ref{angular} we discuss the effect of an angular momentum term. 
Section \ref{degeneracy} is devoted to the study of the ground--states and 
excitations at the degeneracy points, while our conclusions are in Section 
\ref{conclusions}. 
 
\section{Single--particle Hamiltonian and Landau levels} 
\label{intro}

In this Section we first introduce the single--particle Hamiltonian
of a single--component gas in a constant magnetic field in order to set the notation 
for the following results. We then consider a two--component gas 
characterized by two internal degrees of freedom: these two components provide 
a pseudospin degree of freedom (hereafter denoted simply as spin). We 
analyze the general properties of its spectrum when a non-Abelian $SU(2)$ gauge potential independent on the position, 
thus characterized by a constant Wilson loop \cite{goldman09}, is added to a constant magnetic field 
(equal for both the spins).

We show that there are only three classes of quadratic Hamiltonians, 
describing a single particle in such effective $U(2)$ gauge potential, giving rise 
to an analytically defined Landau level structure (with eigenstates 
expressed as a finite linear combination of the eigenstates of a particle 
in a magnetic field). 
The first class corresponds to an Abelian $U(1) \times U(1)$ gauge potential 
and it refers to uncoupled spin states, whereas the other two 
are characterized by a truly non-Abelian $U(2)$ gauge potential and 
correspond to different kinds of Jaynes-Cummings models. 

\subsection{$U(1)$ gauge potentials}
\label{secU1}

We first remind the standard case of a spinless atom 
moving on a plane and subjected to a (fictitious) constant magnetic field 
\cite{yoshioka}. This case corresponds to a $U(1)$ gauge potential: the Hamiltonian reads
\begin{equation}\label{ham_single}
H=\left( p_x+A_x\right)^2+\left( p_y+A_y\right)^2
\end{equation}
(with units such that $\hbar = 1$ and $m=1/2$). 
The vector potential $\vec A$ in the symmetric gauge is 
\begin{equation}
A_x= -\frac{B}{2}y \,,\qquad A_y=\frac{B}{2}x.
\end{equation}
Introducing the complex coordinate $z\equiv x-iy$, we define the 
operators
\begin{equation} \label{algebra} 
D=\frac{1}{2\sqrt{2B}}\left(B z + 4 \partial_{\bar z} \right)
\end{equation}
and 
\begin{equation}
L=\frac{1}{2\sqrt{2B}}\left(B z -4 \partial_{\bar{z}} \right).
\end{equation}
These operators obey the commutation rules
\begin{equation} \label{comm}
\left[ D,D^\dag\right]=1\,,\qquad \left[L^\dag,L \right]=1\,,\qquad 
\left[ L,D \right]=\left[ L,D^\dag \right]=0.  
\end{equation}
$D$ and $D^\dag$ can be considered as ladder operators and allow us 
to express the Hamiltonian (\ref{ham_single}) 
as $H=2B\left(D^\dag D + 1/2\right)$. 
The operators $L$ and $L^\dag$ enter the definition of the angular momentum of 
the particle through the relation 
\begin{equation}
L_z = D^\dag D-L L^\dag
\end{equation}
such that $L$ and $L^\dag$ respectively decrease and increase $L_z$. 
Each energy eigenstate is degenerate with respect to the angular momentum,
therefore it is possible to characterize the usual Landau levels 
by the index $n=D^\dag D$. 

In order to maintain the angular momentum degeneracy of the eigenstates, 
a generic Hamiltonian $H$ for a spinless atom have 
to be independent of the operators $L$ and $L^\dag$. 
Imposing this Hamiltonian to be at most quadratic in the momentum 
we obtain that $H$ can be written as a function of the operators $D$ and $D^\dag$ as
\begin{equation} \label{spinless}
\frac{H}{E'}= D^\dag D + aD^\dag + a^* D + bD^{\dag 2} + b^*D^2 + \mathcal{K'}
\end{equation}
where $E'$ is an overall energy scale, $\mathcal{K'}$ is a real parameter, whereas 
$a$ and $b$ are complex coefficients. 
It is easy to show that the Hamiltonian (\ref{spinless}) 
can be recast in the following form
\begin{equation} \label{spinless2}
H=E \, \Gamma^\dag \Gamma +\mathcal{K}
\end{equation}
where the operators $\Gamma$ is defined as
\begin{equation} \label{aff1}
\Gamma = \frac{1}{2}\left(\alpha + \frac{1}{\alpha^*} \right) D+\frac{1}{2}\left(\alpha - \frac{1}{\alpha^*} \right)D^\dag+\beta 
\end{equation}
in order to satisfy the commutation relation $\left[\Gamma,\Gamma^\dag \right]=1$, with the complex parameters $\alpha$ and $\beta$ related to the coefficients in (\ref{spinless}). The operators $\Gamma,\Gamma^\dag$ allow to express the Hamiltonian in the simple quadratic form (\ref{spinless2}), 
which makes evident the Landau level structure of the 
single--particle problem. 
It is interesting to notice that the mapping from $D,D^\dag$ 
to $\Gamma,\Gamma^\dag$ corresponds to an affine transformation 
of the space coordinates of the kind:
\begin{equation} \label{affine}
z \to z' = \alpha x -\frac{i}{\alpha^*}y + \sqrt{\frac{8}{B}}\beta = 
\frac{1}{2}\left(\alpha + \frac{1}{\alpha^*} \right)z+\frac{1}{2}
\left(\alpha - \frac{1}{\alpha^*} \right)\bar{z}+\sqrt{\frac{8}{B}}\beta.
\end{equation}

\subsection{$U(1) \times U(1)$ gauge potentials} 
\label{secU1xU1}

We consider now a system of (non-interacting) 
atoms characterized by a pseudospin degree of freedom 
that can assume the eigenstates $\left|\uparrow\right\rangle$ and 
$\left|\downarrow\right\rangle$ on the $\hat z$ direction. 
Therefore, in the above description of the single--particle system, 
it is necessary to introduce also the Pauli matrices $\vec \sigma$ 
to complete the observable algebra generated by $D, D^\dag, L$ and $L^\dag$. 
The Pauli matrices $\sigma_x$ and $\sigma_y$ couple the two spin components, 
whereas $\sigma_z$ and the $2 \times 2$ identity matrix $\mathbb{I}$ describe 
particles whose states 
$\left|\uparrow\right\rangle$ and $\left|\downarrow\right\rangle$ are decoupled. 
More generally, to describe a system characterized by a 
$U(1) \times U(1)$ symmetry corresponding to two decoupled states in the same 
magnetic field B, 
the Hamiltonian must be a function of the operators 
$D,D^\dag,\mathbb{I}$ and a single linear combination of Pauli matrices, 
$\hat k \cdot \vec \sigma$, that we can relabel as $\sigma_z$ without loss of generality.

This generic problem is defined by the Hamiltonian
\begin{equation} \label{U1xU1g}
H=\left(a_z\sigma_z + a_0 \right)D^\dag +  \left(a_z^*\sigma_z + a^*_0 \right)D+  \left(b_z \sigma_z + b_0 \right)D^{\dag 2} +  \left(d^*_z \sigma + b^*_0 \right)D^2 
+ \left( h'_z\sigma_z + h'_0\right)D^\dag D + \mathcal{M'}\sigma_z.
\end{equation}
Such Hamiltonian describes a particle in the uniform magnetic field $B$ and it is the most general one which is quadratic in the momentum, fulfils the U(1)$\times$U(1) gauge symmetry and is independent on the angular momentum. 
It constitutes a simple generalization of the Hamiltonian (\ref{spinless}) 
to two non--interacting spin components, 
thus it can be solved by implementing two different coordinate 
transformations of the kind (\ref{affine}) for the two inner states. 
Using the unitary transformation 
$U=e^{-i\sigma_z\left(i \alpha \bar{z} - i \alpha^* z\right) }$, 
the operators $D$ and $D^\dag$ become 
\begin{equation}
D\to U^\dag D U = D + \sqrt{\frac{2}{B}}\,\alpha\sigma_z\,,\qquad D^\dag \to U^\dag D^\dag U = D^\dag + \sqrt{\frac{2}{B}}\,\alpha^*\sigma_z.
\end{equation}
This unitary transformation, 
combined with the real space affine transformations (\ref{affine}), 
allows us to define the operator
\begin{equation}
\Gamma=\left( \alpha_0 + \alpha_1 \sigma_z\right)   D + \left( \beta_0 +  \beta_1 \sigma_z\right)  D^\dag + \gamma_0 + \gamma_1 \sigma_z  
\end{equation}
with the parameters chosen in order to satisfy the commutation relation $[\Gamma,\Gamma^\dag]=1$. Apart from constant terms, the Hamiltonian (\ref{U1xU1g}) 
can be rewritten as
\begin{equation} \label{U1xU1}
H_a=\Gamma^\dag \Gamma \left(h_0 \mathbb{I } + h_z \sigma_z\right) + \mathcal{M}\sigma_z 
\end{equation}
with $h_0$, $h_z$ and $\mathcal{M}$ real parameters. 
Both the spin $\sigma_z$ and the angular momentum $L_z$ 
are proper quantum numbers and $\mathcal{M}$ constitutes a Zeeman term. 
Thus, the spectrum of the Hamiltonian is characterized 
by the presence of Landau levels, degenerate with respect to 
$L_z$ and labelled by both $\sigma_z$ and $n=\Gamma^\dag \Gamma$: 
if $h_z$ and $\mathcal{M}$ go to zero, all the Landau levels result degenerate 
also with respect to the spin.

We conclude this Section by observing that every 
single--particle Hamiltonian of the kind (\ref{U1xU1g}) 
can be recast in the form (\ref{U1xU1}) with suitable transformations. 
Equation (\ref{U1xU1}) makes evident the existence of 
a quantum number $n=\Gamma^\dag \Gamma$ 
which defines the Landau level structure. 
The main characteristic of the Hamiltonian $H_a$ is the fact that 
it does not depend on $\sigma_x$ and $\sigma_y$, and 
therefore it corresponds to a gauge symmetry $U(1) \times U(1)$: 
we will refer to this case as the {\em Abelian limit} of the more general $U(2)$ 
we are going to introduce in the following.

\subsection{$U(2)$ gauge potentials} 
\label{U2}

In this Section we consider particles subjected both to an artificial magnetic field 
and to a general artificial $SU(2)$ homogeneous non--Abelian gauge potential 
(simulating a general spin--orbit coupling). 
Our goal is to characterize the conditions under which the non-Abelian 
gauge potential preserves the Landau levels. 
More precisely, we want to classify the single--particle Hamiltonians 
with a general homogeneous non--Abelian term added to a magnetic field 
(equal for both spins) such that the following properties are satisfied:
\begin{enumerate}
\renewcommand{\theenumi}{$\mathcal{P}$\arabic{enumi}}
\renewcommand{\labelenumi}{ \theenumi :}
\item Their energy spectrum presents a Landau level structure. \label{cond1}
\item Every Landau level is degenerate with respect to the angular momentum 
$L_z$. \label{cond2} 
\item In the Abelian limit, the Landau levels become degenerate with respect of the spin degree of freedom. \label{cond3} 
\end{enumerate}

The condition \ref{cond1} about the existence of the Landau levels is very general: 
it is indeed known that for a broad class of spin--orbit interactions 
the spectrum of the single--particle Hamiltonian is composed by eigenstates 
expressed as infinite series \cite{zhang}. For simplicity, in the following, 
we restrict our attention to Landau levels 
such that the corresponding wavefunctions can be expressed as a finite sum of terms.

We observe that the condition \ref{cond3} 
is not strictly necessary to obtain a proper Landau level structure, 
however it is necessary to implement the $U(2)$ gauge symmetry 
which will characterize the Hamiltonian we will investigate in the following. 
In general, terms as the ones in $h_z$ and $\mathcal{M}$ in (\ref{U1xU1}) 
break this symmetry and do not satisfy the condition \ref{cond3}. 
Nevertheless they do not spoil the main characteristics of the system we will study in section \ref{oneparticle} and can be easily taken into account in the following analysis.


To obtain the most general Hamiltonians satisfying the conditions 
\ref{cond1} - \ref{cond3}, we use the single--particle algebra defined in Sections 
\ref{secU1} - \ref{secU1xU1}. 
Condition \ref{cond2} implies that the spectrum of the Hamiltonian 
must be degenerate with respect to the angular momentum $L_z$. 
Therefore one obtains $\left[H,L\right]=\left[H,L^\dag\right]=0$ and the Hamiltonian 
does not depend on $L$ and $L^\dag$, but only on $D,D^\dag,\vec{\sigma}$. 
As in the previous case (\ref{U1xU1g}), 
spin--orbit couplings can give rise, in general, 
to terms in the Hamiltonian that are not proportional to $D^\dag D$. 
Therefore it is necessary to introduce a wider class 
of ladder operators $\Gamma$ and $\Gamma^\dag$ 
generalizing the operators $D$ and $D^\dag$ previously introduced 
(see \cite{zhang} for an accurate description based on the Landau gauge). 
Condition \ref{cond1} can be rephrased in terms of these generalized 
ladder operators imposing that there must exist an integral of motion $n$ characterized by
\begin{equation}
\left[H,n \right]=0 \,, 
\qquad \left[ n,\Gamma^\dag\right]=\Gamma^\dag \,, \qquad
\left[ n,\Gamma\right]=-\Gamma \label{eqndd},
\end{equation}
where $\Gamma$ obey the commutation relations 
$\left[\Gamma,\Gamma^\dag\right]=1$ and 
$\left[\Gamma,L\right]=\left[\Gamma,L^\dag \right]=0$, and there exists an eigenstate $\Psi_0$ of $\Gamma$ with eigenvalue $0$: 
$\Gamma \Psi_0 = 0$. $\Gamma$ can be defined as a linear combination of 
$D$, $D^\dag$ and, eventually, some constant terms. 

The role of the integral of motion $n$ is to label the generalized Landau levels obtained from the couplings between the pseudospin states. 
To satisfy (\ref{eqndd}), $n$ 
must be chosen in the form
\begin{equation} \label{eqn}
 n=\Gamma^\dag \Gamma + \vec{c}\cdot\vec{\sigma},
\end{equation}
where $\vec{c}$ is a real vector we want to determine.

We define now the most general single--particle Hamiltonian 
satisfying the previous conditions, 
with the constraint that it can be at most quadratic in the momentum 
$\vec{p}$ (and, therefore, in the operators $D$ and $D^\dag$). 
We can divide the Hamiltonian into two terms: 
the first one, $H_a$, corresponds to the Abelian gauge symmetry $U(1) \times U(1)$ and 
represents the case of uncoupled spin components:
\begin{equation} \label{ha}
H_a=\left( h_0 + h_z\sigma_z\right)D^\dag D + \mathcal{M}_z\sigma_z.
\end{equation}
The second term, $H_{na}$, is instead the non--Abelian contribution $H_{na}$
\begin{multline} \label{hna}
H_{na}=\left(\vec a \cdot \vec \sigma + a_0 \right)D^\dag +  \left(\vec{a^*} \cdot \vec \sigma + a^*_0 \right)D+  \left(\vec b \cdot \vec \sigma + b_0 \right)D^{\dag 2} +  \left(\vec{b^*} \cdot \vec \sigma + b^*_0 \right)D^2 
+ \left( h_x\sigma_x+h_y\sigma_y\right)D^\dag D + \mathcal{M}_x\sigma_x + \mathcal{M}_y \sigma_y
\end{multline}
where $\vec{\mathcal{M}}$ is a real vector, $h^\mu$ 
is a real vector with spatial part $\vec h$, and $a^\mu$, $b^\mu$ are 
complex vectors with spatial parts $\vec{a}$ and $\vec b$. 
The total Hamiltonian (apart from constant terms) 
is given by $H=H_a + H_{na}$. 

We observe that, if the following conditions are satisfied:
\begin{itemize}
 \item $\vec a = e^{i\theta_a}\vec{a_R}$, with  $\vec{a_R}$ a real vector and $\theta_a$ a real constant;
 \item $\vec b = e^{i\theta_b}\vec{b_R}$, with  $\vec{b_R}$ a real vector and $\theta_b$ a real constant;
 \item all the vectors $\vec{\mathcal{M}},\vec{h},\vec{b_R}$ and $\vec{a_R}$ are parallel to each other along the direction $\hat k$;
\end{itemize}
then the Hamiltonian can be reduced to the $U(1)\times U(1)$ case 
(\ref{U1xU1g}) previously studied through proper transformations. 
In this case we have shown that there is a Landau level operator 
$n=\Gamma^\dag \Gamma$ commuting with the Hamiltonian. 
The previous conditions imply that the system is not 
characterized by a proper $U(2)$ gauge potential since 
only one effective component of the spin, $\hat k \cdot \vec \sigma$, 
enters the Hamiltonian. 

In the generic $U(2)$ case with arbitrary vectors $\vec{a},\vec{b}$, 
it is impossible to gauge away all the terms in $\vec{a} \cdot \vec{\sigma}$. 
It is therefore convenient to search for an integral of motion $n$ 
defined in terms of the $D$ operators as 
$n=D^\dag D + \vec c \cdot \vec \sigma$; 
in order to satisfy the relation $[H,n]=0$, 
we obtain from the commutation rules the conditions
\begin{eqnarray}
 &\epsilon^{lmn}h_lc_m = \epsilon^{lmn}\mathcal{M}_lc_m = 0 \label{eqzc}\\
&-a_n + 2i\epsilon^{lmn}a_lc_m =0\,,\qquad a^*_n + 2i\epsilon^{lmn}a^*_lc_m =0\,,\qquad  a_0 = 0  \label{eqac} \\
&-2b_n + 2i\epsilon^{lmn}b_lc_m =0\,,\qquad 2b^*_n + 2i\epsilon^{lmn}b^*_lc_m =0\,,\qquad  b_0 = 0  \label{eqbc} 
\end{eqnarray}
Equation (\ref{eqzc}) requires the vectors 
$\vec h, \vec{\mathcal{M}}$ and $\vec c$ to be parallel; 
without loss of generality, we can impose them to be in the 
$\hat z$ direction with an appropriate spin rotation and 
$\vec c$ assumes the form $(0,0,c_z)$. 
The equations (\ref{eqac}) and (\ref{eqbc}) 
are not compatible in general, unless either $\vec a$ or $\vec b$ is zero. 
Imposing in (\ref{eqac}) 
$a_0=0$ and $b_0=0$, by using the transformation (\ref{aff1}), 
Equation (\ref{eqac}) can be recast in the form
\begin{eqnarray}
\Im\left( \vec a\right) &=& 2 \Re\left( \vec{a}\right)  \times \vec{c} \label{Ima}\\
\Re\left( \vec a\right) &=&-2 \Im\left( \vec{a}\right) \times \vec{c} \label{Rea}.
\end{eqnarray}
The previous equations state that the real and imaginary parts of 
$\vec a$ have to be orthogonal to each other and orthogonal to $\vec{c}$. 
Therefore the condition (\ref{eqndd}) implies, 
in the case $\vec b=0$, that $\vec a$ must lie in the $xy$ plane and, 
moreover, that $\left|c_z\right| = 1/2$ 
in order to satisfy (\ref{Ima}) and (\ref{Rea}). Therefore 
we can choose a proper spin basis in which 
\begin{equation}
\vec{a}=\left( iq/2,q/2,0\right) \,,\qquad \vec{c}=\left(0,0,1/2 \right). 
\end{equation}
Similarly one can consider the case in which $\vec{a}=0$: to satisfy equation (\ref{eqbc}), $\vec{b}$ must lie in the system plane and $\Re({\vec b}) \perp \Im({\vec b})$ with $\left|c_z\right| = 1$ (which is incompatible with the previous case).

So far we considered conditions \ref{cond1} and \ref{cond2} 
and we obtained, in the general $U(2)$ case, 
that there are two possible Hamiltonian classes 
defined by $\vec{b}=0$ and $\vec{a}=0$. The transformation 
(\ref{aff1}) required to obtain $a_0=0$ and $b_0=0$ 
implies that the Landau level operator $n$ 
has the general form (\ref{eqn}) 
in terms of the operators $\Gamma$ and $\Gamma^\dag$. 
For the sake of simplicity, hereafter 
we restrict ourself to the case $\Gamma=D$ and $\Gamma^\dag=D^\dag$, 
since all the other cases can be studied using the 
coordinate transformation (\ref{affine}). Therefore 
we will deal with generalized Landau levels expressed as a 
finite sum of the eigenstates of $D^\dag D$. 
In this case we are reducing the previous Hamiltonians to the following classes:
\begin{itemize}
\item The Jaynes--Cummings class, obtained by imposing $\vec b =0$, with 
Hamiltonian 
\begin{equation} \label{class1}
H=\left( E + h_z\sigma_z\right)D^\dag D + \mathcal{M}_z\sigma_z + \mathcal{K} - i q \sigma_+ D + i q \sigma_- D^\dag
\end{equation}
where $\sigma_\pm = \sigma_x \pm i \sigma_y$. 
In the limit $h_z,\mathcal{M}_z \to 0$ this Hamiltonian 
satisfies also the condition \ref{cond3} and 
it is characterized by a full $U(2)$ gauge symmetry. 
Moreover, it can be shown that the Hamiltonian (\ref{class1}) 
is gauge equivalent to both a pure Rashba spin--orbit coupling and 
a pure Dresselhaus interaction. 
This case will be extensively discussed in the next Sections 
where we will show that it can be described in terms 
of a minimal coupling with a non--Abelian gauge potential.
\item The two--photon Jaynes--Cummings class, 
obtained by imposing $\vec a =0$, corresponding to
\begin{equation}
H=\left( E + h_z\sigma_z\right)D^\dag D + \mathcal{M}_z
\sigma_z + \mathcal{K} - i q \sigma_+ D^2 + i q \sigma_- D^{\dag 2}.
\end{equation}
This Hamiltonian cannot be described by a quadratic minimal coupling 
with a non--Abelian gauge potential 
since it presents the product between quadratic terms in $D$ and $D^\dag$ 
and the $\sigma$ matrices. However it can be exactly solved 
\cite{gerry,brihaye}, showing a Landau level structure. 
Also in this case the condition \ref{cond3} is satisfied in the limit 
$h_z,\mathcal{M}_z \to 0$.
\end{itemize}

Summarizing, up to transformations of the spin basis, 
there are only three classes of Hamiltonians, 
quadratic in the momentum, that satisfy the conditions 
\ref{cond1} and \ref{cond2} and can be described by generic ladder 
operators $\Gamma$ and $\Gamma^\dag$. 
The first one is the Abelian class with a $U(1)\times U(1)$ 
gauge symmetry (\ref{U1xU1}). 
The other two are characterized by a full $U(2)$ gauge potential and 
correspond to different Jaynes--Cummings models. 
In particular we will restrict to the case 
in which $\Gamma$ and $\Gamma^\dag$ correspond to 
$D$ and $D^\dag$ and we will focus, 
in the following Sections, on the class (\ref{class1}) 
in the limit of $U(2)$ gauge symmetry, 
fulfilling also the condition \ref{cond3}.

\section{Engineering the Non--Abelian Gauge Potential} 
\label{sectripod}

In this Section we discuss the physical implementation in a rotating 
tripod system of the non-Abelian gauge potential 
\begin{equation} \label{eq1}
A_x=q\sigma_x-\frac{B}{2}y\mathbb{I} \; , 
\qquad A_y=q\sigma_y+\frac{B}{2}x\mathbb{I}\,
\end{equation}
(the identity matrix $\mathbb{I}$ will be dropped in the following). 
The corresponding Hamiltonian is
\begin{equation}\label{ham}
H=\left( p_x+q\sigma_x-\frac{B}{2}y\right)^2+\left( p_y+q\sigma_y+\frac{B}{2}x\right)^2.
\end{equation}
As discussed in the previous Section, the vector potential (\ref{eq1}) 
is representative of a much more general class 
of single--particle systems (\ref{class1}) characterized by the properties 
\ref{cond1} - \ref{cond3}. Moreover the non--Abelian term 
of (\ref{eq1}) mimics the effect of a spin--orbit coupling and 
it can be shown to be gauge equivalent both 
to the Dresselhaus and to the Rashba coupling.

The vector potential (\ref{eq1}) is constituted by a $SU(2)$ 
term proportional to the parameter $q$, quantifying the strength of the 
non--Abelian term, and by the magnetic contribution: 
$\vec A$ describes a proper $U(2)$ potential whose 
total effective magnetic field is
\begin{equation} \label{eq1a}
\mathcal{B}=\nabla \times \vec A + i \vec A\times\vec A= \begin{pmatrix} B-2q^2 &0 \\ 0 &B+2q^2 \end{pmatrix} 
\end{equation} 
where $\mathcal{B}$ is proportional to the commutator of the 
covariant derivatives. It is important to notice that $\mathcal{B}$ 
does not depend on the position so that the system is characterized 
by a translationally invariant Wilson loop 
as in the cases analyzed in \cite{goldman09}. 

It is well known that the effect of a constant magnetic field 
can be reproduced in a rotating frame thanks to the Coriolis force 
(see, for example, \cite{cooper08}); at variance, 
the $SU(2)$ contribution can be obtained through proper optical 
couplings in a system of atoms showing 
quasi-degenerate ground--states as described in \cite{ruse05,santos07}. 
Therefore, to engineer the effective gauge potential 
$\vec A (q)$ (\ref{eq1}), we will consider a rotating system 
of the so-called tripod atoms whose coupling 
is described in \cite{ruse05} and depends on the Rabi frequencies $\Omega_i$
\begin{equation} \label{rabif}
\Omega_1 = \varOmega \sin\left( \theta\right)\cos\left(\phi\right)e^{iS_1}\,,\quad \Omega_2 = \varOmega \sin\left( \theta\right)\sin\left(\phi\right)e^{iS_2}\,,\quad \Omega_3=\varOmega\cos\left(\theta\right)e^{iS_3},  
\end{equation}
where $S_1$ and $S_2$ are functions of the position and of the parameter $q$, 
while the angles $\phi$ and $\theta$ and $S_3$ are chosen constants. 
These frequencies describe the couplings of 
three quasi-degenerate ground--states, 
characterized by different hyperfine levels, with an excited state. 
This interaction give rise to two different dark states 
whose dynamics is described by the effective Hamiltonian (\ref{ham}) 
with vector potential given by (\ref{eq1}). 
Such dark states constitute the different pseudospin component 
$\left| \uparrow \right\rangle$, $\left| \downarrow \right\rangle$ 
coupled by the non--Abelian term in $H$.

In the following we determine the dependence of the frequencies 
$\Omega_i$ on $q$ to obtain, through a proper gauge transformation, 
the Hamiltonian (\ref{ham}) for the atomic gas in a rotating frame. 

Let us consider first a system of tripod atoms in an inertial 
frame of reference characterized by a non--Abelian gauge potential 
$\vec{\tilde{A}}$, a scalar potential $V_{\rm rot} \equiv \Phi(\vec{\tilde{A}}) + V$ 
as the one described in \cite{ruse05} and a harmonic confining potential 
$\omega r^2 / 4$ where $r^2 \equiv x^2+y^2$ (notice that we are working in units 
in which $m=1/2$). The corresponding Hamiltonian reads 
\begin{equation}
 H_{\rm IF}=\left( \vec{p}+\vec{\tilde{A}}\right) ^2 + 
\frac{1}{4}\omega^2 r^2  + V_{\rm rot}.
\end{equation}
Once the whole system is put in rotation with angular velocity $\Omega$, the Hamiltonian in the rotating frame of reference reads \cite{lu07}
\begin{equation}
H_{\rm rot}=\left( \vec{p}+\vec{\tilde{A}} \right) ^2 + \frac{1}{4}\omega^2 r^2 +\Omega \mathbb{L}_z  + V_{\rm rot},
\end{equation}
where we introduced the gauge invariant angular momentum
\begin{equation}
\mathbb{L}= \vec r \times \left(\vec p + \vec{\tilde{A}} \right)
\end{equation}
and all the coordinates are now considered in the rotating frame.
It is useful to rewrite $H_{\rm rot}$ introducing the gauge potential
\begin{equation} \label{eq2}
A_x= \tilde A_x -\frac{B}{2}y\,,\qquad A_y = \tilde A_y +\frac{B}{2}x 
\end{equation}
where we imposed $B=\omega$. We obtain
\begin{equation} \label{eqH2}
 H_{\rm rot}=\left( \vec{p}+\vec{A}\right) ^2 -  \Delta  \mathbb{L}_z  + V_{\rm rot} 
\end{equation}
with $\Delta = \omega - \Omega$.

Our aim is to identify the correct family of Rabi frequencies, 
$V_{\rm rot}$ and gauge transformations such that the Hamiltonian $H_{\rm rot}$ 
can be cast in the form
\begin{equation} \label{eqH}
 H_L=\left( \vec{p}+\vec{A}\right) ^2 -  \Delta  L_z  
\end{equation}
with $\vec{A}$ given by (\ref{eq1}) and 
$L_z=\vec r \times \vec p$ being the usual angular momentum 
in the rotating frame. 
As we will show in section \ref{angular}, $H_L$ can be exactly solved and, 
in the limit $\Omega \to \omega$, it becomes the Hamiltonian (\ref{ham}). In particular we need to have
\begin{eqnarray}
\vec{\tilde{A}} &=& \left(q\sigma_x,q\sigma_y\right), \\
V_{\rm rot} &=& q \Delta (x\sigma_y - y\sigma_x ).
\end{eqnarray}
In order to obtain, in the rotating frame, the potential 
$\vec{A}$ in (\ref{eq1}) 
starting from the potential $\mathcal{A}$ 
\begin{eqnarray}
 \mathcal{A}_{11} &=& -\cos^2 \phi \nabla S_{23}-\sin^2\phi \nabla S_{13} \label{a11} \\
 \mathcal{A}_{22} &=& -\cos^2 \theta \left( \cos^2\phi \nabla S_{13}+\sin^2\phi \nabla S_{23}\right)  \label{a22}\\
 \mathcal{A}_{12} &=& -\cos \theta \left( \frac{1}{2} \sin 2\phi \nabla S_{12}-i\nabla\phi\right)  \label{a12}
\end{eqnarray}
given in \cite{ruse05}, 
we need a suitable unitary gauge transformation $O(\vec r)$ \cite{note1}. 
In particular the field transforms as $\mathcal{A} \to O  \mathcal{A} O^\dag -i O \nabla O^\dag$ and thus we must have
\begin{equation}
 O \vec{\mathcal{A}} O^\dag -i O \vec{\nabla} O^\dag = \vec{\tilde{A}} = \left(q \sigma_x, q \sigma_y\right). 
\end{equation}

From the definition of $\mathcal{A}$ one can see that, 
choosing a constant $\phi$, it is not possible 
to obtain $\mathcal{A}_y \propto \sigma_y$, 
but it is possible to check that we can obtain 
$\mathcal{A}_y = k \mathbb{I} - q\sigma_z$ and 
$\mathcal{A}_x = q\sigma_x$ for a suitable choice of the parameters 
as functions of $q$. Therefore the gauge transformation we apply is
\begin{equation} \label{gauge}
\Psi \to O \Psi \,,\qquad {\rm with} \; O=e^{iky-i\frac{\pi}{4}\sigma_x}
\end{equation}
with $k$ to be defined in the following. In this way we obtain
\begin{equation} \label{eqa1}
\mathcal{A}_x = q \sigma_x \xrightarrow{O} \tilde A_x = q\sigma_x \,,\qquad \mathcal{A}_y = k \mathbb{I} - q\sigma_z \xrightarrow{O} \tilde A_y = q\sigma_y.
\end{equation}
We have also to consider that the scalar potential in \cite{ruse05} 
is affected by $O$ as $O\left( V+\Phi \right)O^\dag$: then, 
in order to obtain (\ref{eqH}) out of (\ref{eqH2}), we need to have
\begin{equation} 
O\left( V+\Phi \right)O^\dag = V_{\rm rot} = \Delta \vec r \times 
\vec{\tilde{A}}
\end{equation}
and then
\begin{equation} \label{eqa2}
V+\Phi = -q\Delta\left( y \sigma_x + x \sigma_z\right) .
\end{equation}

We are now in position to find the suitable parameters 
satisfying Equations (\ref{eqa1}) and (\ref{eqa2}). 
First of all we impose $\phi=\pi/4$ and $S_3={\rm cost}$: then, 
from the definition of $\mathcal{A}$ we obtain that
\begin{eqnarray}
\partial_x \left( S_{1} + S_{2}\right)  =  0 \\
 -\cos \theta \partial_x \left( S_1 - S_2 \right) =2q \\
2k - 2q = -\partial_y \left( S_{1} + S_{2}\right) \\
2q + 2k =-\cos^2 \theta \partial_y \left( S_{1} + S_{2}\right).
\end{eqnarray}
A possible solution is given by
\begin{equation}
S_1=\lambda\left( x + y\right) \,, \qquad S_2=\lambda\left( -x + y\right) 
\end{equation}
with $\lambda = -q/\cos\theta$. From the last two equations we obtain
\begin{equation}
\cos^2 \theta-2\cos \theta -1 = 0 \quad \Rightarrow  \quad \cos \theta = 1 - \sqrt{2}.
\end{equation}
It follows that 
\begin{equation} \label{eqk}
\lambda = -\frac{q}{\cos\theta}=\frac{q}{\sqrt{2}-1} \, , \qquad k = \frac{1+\cos^2\theta}{2\cos\theta}\,q=-\sqrt{2}q.
\end{equation}
The corresponding Rabi frequencies are
\begin{equation} \label{eqrabi}
\Omega_1 = \varOmega' e^{iq\frac{x+y}{\sqrt{2}-1}}\,,\quad \Omega_2 = \varOmega' e^{iq\frac{-x+y}{\sqrt{2}-1}}\,,\quad \Omega_3=-\varOmega' \sqrt{\sqrt{2}-1}e^{iS_3}\simeq -0.64 \varOmega' e^{iS_3}
\end{equation}
with $\varOmega'$ and $S_3$ arbitrary, so that the right $\vec{\tilde{A}}$ 
are obtained after the gauge transformation.

Let us consider now the scalar potentials; 
imposing $\phi = \pi/4$ and 
$\cos\theta=1-\sqrt{2}$, we find from \cite{ruse05} and from (\ref{eqa2})
\begin{eqnarray}
V_{11}+\Phi_{11}&=&\frac{V_1+V_2}{2}+\lambda^2\sin^2\theta=  -q \Delta x \\
V_{22}+\Phi_{22}&=&\frac{V_1+V_2}{2}\cos^2\theta+V_3\sin^2\theta + \lambda^2\cos^2\theta\sin^2\theta = q \Delta x\\
V_{12}+\Phi_{12}&=&\frac{V_1-V_2}{2}\cos\theta=-q \Delta y.
\end{eqnarray}
The solution is given by
\begin{eqnarray}
V_1&=&-q \Delta x - \lambda \Delta y - \lambda^2 \sin^2\theta  \label{v1}\\
V_2&=&-q \Delta x + \lambda \Delta y - \lambda^2 \sin^2\theta  \label{v2}\\
V_3&=& \sqrt{2} q \Delta x 
\end{eqnarray}
with $\lambda$ given by (\ref{eqk}).

This choice for the scalar potentials $V_i$ completes the set of parameters 
(\ref{eqrabi}) needed to obtain the Hamiltonian (\ref{eqH}) 
in the rotating frame. We notice that it is 
possible to modify the previous derivation of the scalar potential 
in order to obtain a Zeeman splitting reproducing 
the term proportional to $\mathcal{M}_z$ in (\ref{class1}).

We conclude this Section observing that it is impossible 
to obtain the gauge potential $\vec A$ defined in Equation (\ref{eq1}) 
(or a gauge equivalent version) using only the gauge potential 
$\vec{\mathcal{A}}$ (\ref{a11},\ref{a22},\ref{a12}) defined in \cite{ruse05} 
without introducing other physical elements such as the rotation of the system. 
Indeed, applying the gauge transformation $O^\dag$ (\ref{gauge}) 
we can express $\vec A$ as
\begin{equation} \label{eq1app}
A_x=-\frac{B}{2}y + q\sigma_x\,,\qquad A_y=\frac{B}{2}x-q\sigma_z + k.
\end{equation}
This form of the potential $\vec A$ is real and does not depend on 
$\sigma_y$. Therefore, imposing $\vec{\mathcal{A}}=\vec{A}$ and 
considering Equation (\ref{a12}), we obtain that the parameter 
$\phi$ entering the definition of the Rabi frequencies (\ref{rabif}) 
must be independent on the position, otherwise an imaginary term 
proportional to $\nabla \phi$ would appear.
Let us consider now the term $\vec{\mathcal{A}}_{11}$: 
from the equation (\ref{a11}) one obtains that 
$\partial_y \mathcal{A}_{11,x} = \partial_x \mathcal{A}_{11,y}$. 
However this is not the case for $A_{11,x}$ and $A_{11,y}$ 
in (\ref{eq1app}) unless $B=0$. 
Therefore to obtain the term proportional to the magnetic field $B$ 
one needs either more complicated multipod schemes or 
an additional physical mechanism (in our analysis the rotation) 
besides the construction of the non-Abelian gauge potentials for tripod atoms.

\section{Landau level spectrum} 
\label{oneparticle}

In this Section we discuss the diagonalization of the 
single--particle Hamiltonian in the non--Abelian gauge potential 
described in the previous Section: 
for the sake of simplicity, we will begin our analysis 
studying the Hamiltonian (\ref{ham}), corresponding 
to the limit  $\Omega \to \omega$, 
which is necessary to satisfy the condition \ref{cond2} in section \ref{U2}. 
In the next Section, we will address the case in which also 
a term linear in the angular momentum (\ref{eqH}) is present. 
Two--body interactions will be introduced in Sections \ref{twobody} and 
\ref{gener}.

As discussed in Section \ref{intro}, 
the Hamiltonian (\ref{ham}) can be decomposed into two terms: 
the Abelian one, $H_a$, and the non--Abelian one, $H_{na}$. One has 
\begin{eqnarray} \label{eqhz}
H_a&=&2q^2+B+\frac{1}{4}\left( B\bar z-4\partial_z\right) \left(Bz+4\partial_{\bar z }\right) = 2q^2+B+\frac{1}{4}d^\dag d\\
H_{na}&=&q\begin{pmatrix}0&-iBz -4i\partial_{\bar z} \\ iB\bar z -4i\partial_z & 0\end{pmatrix} = q \begin{pmatrix}0&-id\\id^\dag&0\end{pmatrix} \label{eqhna}
\end{eqnarray}
where we defined the operator
\begin{equation} \label{eqD}
d=Bz+4\partial_{\bar z }=2\sqrt{2B}D
\end{equation}
proportional to the previously defined operator $D$ so that the relations 
$\left[ d,z\right] =0$ and $\left[ d,d^\dag\right]=8B$ hold. 

Introducing the standard gaussian wavefunction 
$\psi_0=e^{-\frac{Bz\bar{z}}{4}}$, one has $d\psi_0=0$ and 
$d^\dag \psi_0=2B\bar z \psi_0$; thus, for $H_a$, 
we obtain the usual Landau level structure of the eigenstates \cite{yoshioka}, 
degenerate with respect to the angular momentum $n-m$:
\begin{equation}
\psi_{n,m} = \frac{i^n  d^{\dag\,n} \left( z^m \psi_0\right)}{\left( 8B\right)^{\frac{n}{2}} \sqrt{n!}   }\propto D^{\dag\, n} L^m \psi_0.
\end{equation}
The corresponding energy levels are
\begin{equation}\label{en}
E_n=2q^2+2B\left( n+\frac{1}{2}\right).
\end{equation}
The eigenvalues of $H_{na}$ are $\lambda^\pm = \pm2q\sqrt{2Bn }$ and 
the corresponding eigenstates $\varphi_{n,m}^\pm$ 
can be expressed in terms of the eigenstates of $H_a$ as 
\begin{equation} \label{eqc}
 \varphi_{n,m}^\pm=\psi_{n-1,m}\left|\uparrow\right\rangle \pm \psi_{n,m}\left|\downarrow\right\rangle
\end{equation}
where the following relation holds for $n\ge 1$:
\begin{equation} \label{eqb}
2\sqrt{2Bn } \psi_{n,m}= id^\dag \psi_{n-1,m}.
\end{equation}

In general the non--Abelian term mixes the $\left( n-1\right) ^{\rm th}$ 
and the $n^{\rm th}$ Landau levels; but there are also uncoupled 
eigenstates $\varphi_{0}=\psi_{0,m}\left|\downarrow\right\rangle$ 
with eigenvalue $\lambda=0$ for every $\psi_{0,m}$ in the lowest Landau level. 
The spectrum of $H_{na}$ is similar to the one 
obtained in the relativistic case typical of the graphene systems 
\cite{neto09}; in particular, these results are analogous to the ones 
obtained in \cite{lewen09} starting from the Dirac equation 
in an anisotropic regime, and we can notice that $H_{na}$ corresponds 
to the known Jaynes---Cummings model, as discussed in Section \ref{intro}.

\begin{figure}[ht]
\centering
\includegraphics[width=8cm]{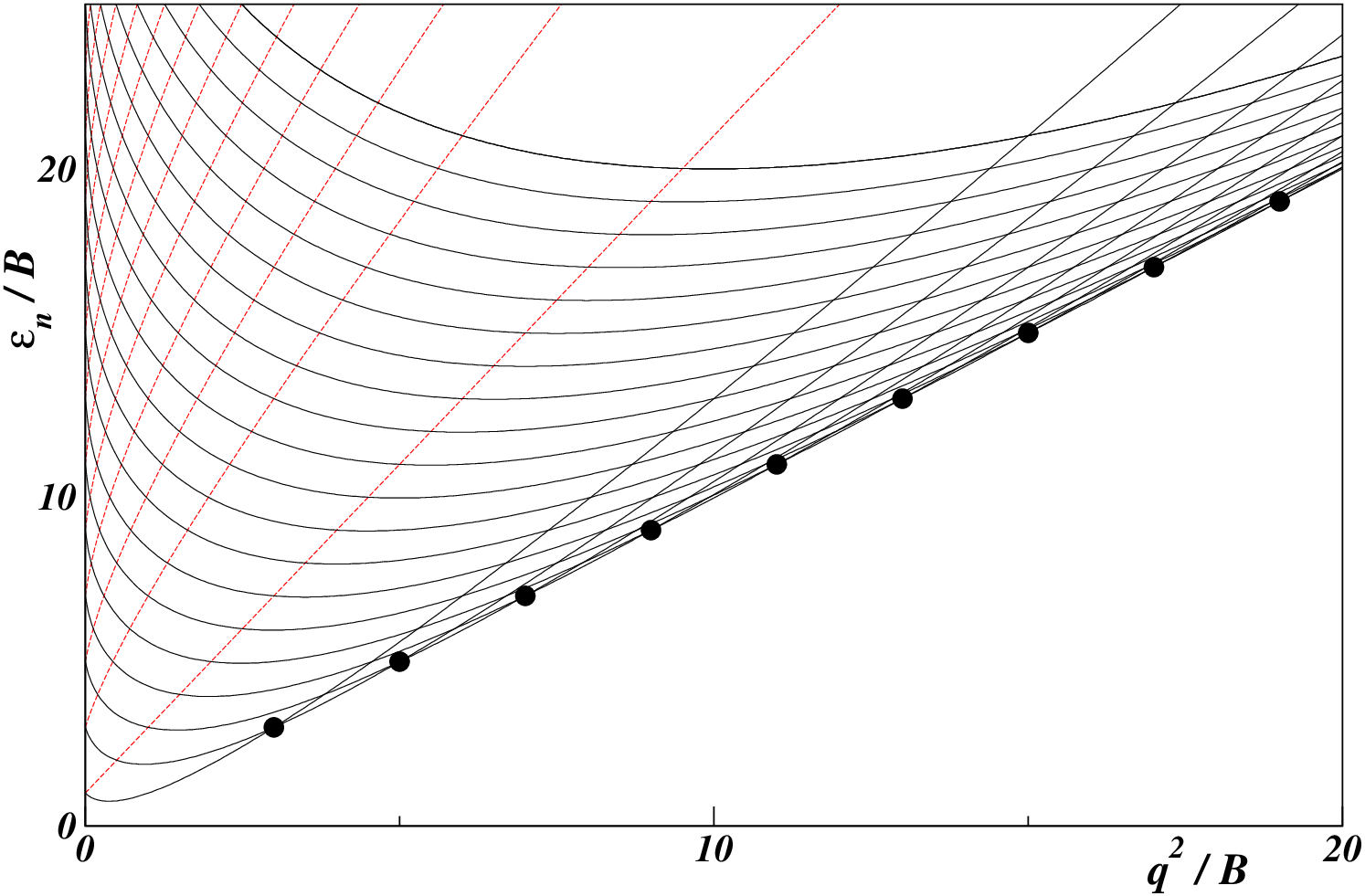}
\caption{Energies of $\chi_{n+1}^-$ (black solid line) and $\chi_n^+$ 
(red dashed line) for $n=0,\dots,10$ as a function of $q^2/B$. 
The crossings of the ground--states occur in the degeneracy 
points $q^2/B=\left(1+2n \right)$, denoted by solid circles.}
\label{fig1}
\end{figure}

We can now diagonalize the whole Hamiltonian 
using as a basis the functions $\varphi_n^\pm$: it is 
\begin{equation}
H\varphi_n^\pm=\left( 2q^2+2Bn\pm 2q \sqrt{2Bn}\right)
\varphi_n^\pm-B\varphi_n^\mp, 
\end{equation}
so that, for $n\ge1$, the Hamiltonian is splitted in blocks $H_n$ of the form
\begin{equation}
\begin{pmatrix}  2q^2+2Bn + 2q \sqrt{2Bn}   & 
-B \\ -B & 2q^2+2Bn - 2q \sqrt{2Bn} 
\end{pmatrix}.
\end{equation}
For the uncoupled states one has 
\begin{equation} \label{uncoupled}
 H \varphi_{0}=\left( B+2q^2\right) \varphi_{0}.
\end{equation}
The eigenvalues of $H$ are therefore
\begin{equation} \label{enlev}
\varepsilon_n^{\pm}=2Bn+2q^2\pm\sqrt{B^2+8q^2Bn } 
\end{equation}
and its (unnormalized) eigenstates are
\begin{equation} \label{eqeig}
\chi_{n,m}^\pm=\left( B + 2q \sqrt{2Bn } \mp \sqrt{B^2+8q^2Bn}\right) \psi_{n-1,m}\left|\uparrow\right\rangle 
+ \left( B - 2q \sqrt{2Bn } \pm \sqrt{B^2+8q^2Bn}\right) \psi_{n,m}\left|\downarrow\right\rangle,
\end{equation}
where we made explicit the angular momentum degeneracy. 
A plot of the eigenvalues of $H$ is presented in Fig. \ref{fig1}.

We notice that the angular momentum is not a good quantum number 
for this states unless we introduce also a spin--$1/2$ component 
defining a total angular momentum $J=L+S$. 
$J$ commutes with both the Hamiltonians (\ref{eq1}) and (\ref{eq2}) 
and $J\chi_{n,m}^\pm = \left(n - m - 1/2 \right) \chi_{n,m}^\pm$. 
Moreover, the wavefunctions $\chi_{n,m}^\pm$ are eigenstates 
for the operator $n=D^\dag D + \sigma_z /2$ defined in (\ref{eqn}): thus, 
they constitutes deformed Landau levels.

Analyzing the spectrum, we can notice that there 
is a correspondence between the usual Landau levels and the states $\chi$'s \cite{trombettoni10}: 
every Landau level is splitted into two parts corresponding 
to the states $\chi_{n-1}^+$ and $\chi_n^-$ and, in the case $q\to 0$, 
their energy becomes approximately $E_{n-1}\pm 4 q^2n$. 
Therefore, for $q\to 0$ one recovers the usual Landau levels structure 
characterized by the (double) spin degeneracy as prescribed by the condition \ref{cond3}. 
The non-Abelian term of the Hamiltonian removes this degeneracy 
through the coupling between the $\left( n-1\right) ^{\rm th}$ level with spin up 
and the $n^{\rm th}$ with spin down. 
The deformed Landau levels for $q>0$ can be defined also considering the Landau gauge; 
in \cite{palmer11} it is shown that, in this case, 
the eigenstates are distinguished by the $Z_2$ 
symmetry obtained by the parity transformation $x \to -x$.

\begin{figure}[ht]
\centering
\includegraphics[width=8cm]{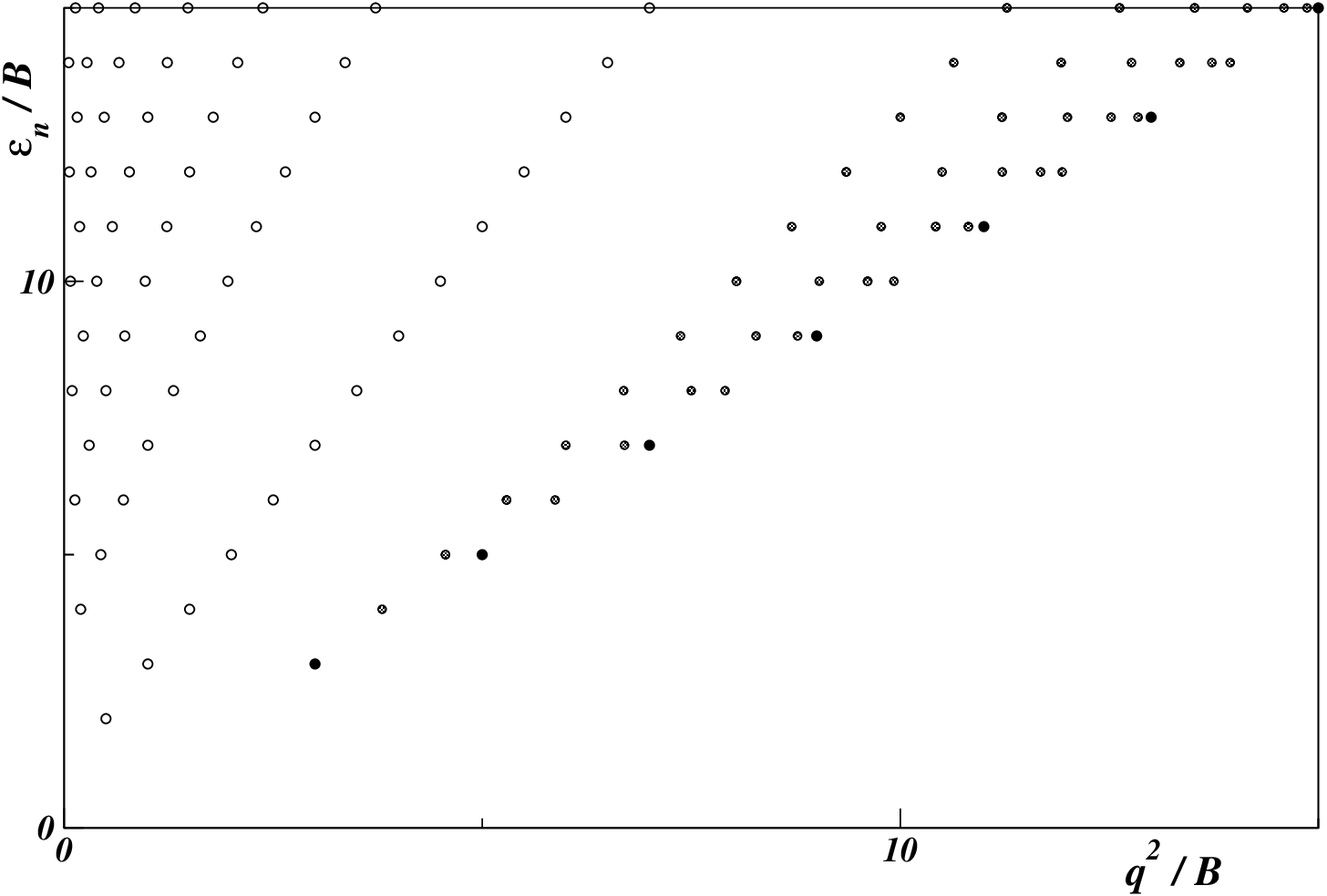}
\caption{Crossing points of the eigenvalues: black (gray) circles indicate 
the crossings between states $\chi_{n+1}^-$ corresponding to the ground--state 
(excited states). Whereas circles denote the crossings of the states $\chi_{n+1}^-$ with 
the states $\chi_n^+$ ($n=0,\dots,10$). The coordinate $q^2/B$ for these degeneracy points are provided in Equations (\ref{crossmm},\ref{crossmp}).}
\label{fig2}
\end{figure}

Varying the value of the parameter $q^2/B$, measuring 
the ratio of the Abelian and non-Abelian
contribution in the gauge potential, 
the eigenvalues $\varepsilon_n^{\pm}$ show an interesting pattern of crossing points 
(see Fig. \ref{fig2}): each pair of eigenstates of the kind $\chi^-_a$ and $\chi^-_b$ 
becomes degenerate for
\begin{equation} \label{crossmm}
\frac{q^2}{B} = \frac{1}{2}\left( a+b + \sqrt{1+4ab}\right) 
\end{equation}
and the energy of the crossing is $\varepsilon_c\left(a,b \right) = \left(a+b \right)B$. 
Instead, a pair of different eigenstates of the kind $\chi^+_a$ and $\chi^-_b$ 
has a crossing only if $a<b$; in this case the degeneracy point is  
\begin{equation} \label{crossmp}
\frac{q^2}{B} = \frac{1}{2}\left( a+b - \sqrt{1+4ab}\right):  
\end{equation}
also for these levels the corresponding energy is 
$\varepsilon_c\left(a,b \right) = \left(a+b \right)B$. 
Therefore all the energy level crossings are characterized 
by an integer energy in units of $B$. 

As shown in Fig. \ref{fig1} the uncoupled state family (\ref{uncoupled}), 
corresponding to $\chi_{0}^+$, is characterized 
by the energy $\varepsilon_0^+ = B+2q^2$, 
which is higher than the energy $\varepsilon_1^- = 2B+2q^2-\sqrt{B^2+8Bq^2}$ 
of $\chi_1^-$. Therefore $\chi_1^-$ is the ground--state family of the system for 
$q^2<3B$ (the general case with $q^2\ge 3B$ is analyzed in section \ref{gener}). 
We can rewrite each state of the family $\chi_1^-$ in the form
\begin{equation} \label{gs}
\chi_1^- = e^{-\frac{B}{4}\left|z\right|^2}\left(  c_{\uparrow,1} 
P \left|\uparrow\right\rangle +2c_{\downarrow,1} B \bar{z} 
P \left|\downarrow\right\rangle - 4c_{\downarrow,1} \partial_z 
P \left|\downarrow\right\rangle\right), 
\end{equation}
where $P$ is a generic polynomial in $z$ and we defined the constants
\begin{eqnarray} \label{constc}
c_{\uparrow,n} &=& B+2q\sqrt{2Bn }+\sqrt{B^2+8q^2Bn } \\
c_{\downarrow,n} &=&i\left({B- 2q \sqrt{2Bn }-\sqrt{B^2+8q^2Bn }}\right)\left(2\sqrt{2Bn}\right)^{-1}. \label{constcb}
\end{eqnarray}
It is also convenient to introduce the operator
\begin{equation} \label{eqG}
\mathcal{G}_1 \equiv  c_{\uparrow,1} \sigma_x + c_{\downarrow,1} d^\dag
\end{equation}
that allows us to map uncoupled states in $\chi_0^+$ in states in $\chi_1^-$ 
\cite{trombettoni10}. 
Thus, we can rewrite (\ref{gs}) using the operator $\mathcal{G}_1$
\begin{equation}
\chi_1^-=\mathcal{G}_1 \left(P(z) e^{-\frac{B}{4}\left|z\right|^2}
\left|\downarrow\right\rangle\right).
\end{equation}

Due to the degeneracy of these ground--states, one can build also 
wavefunctions minimizing supplementary terms in the Hamiltonian (\ref{ham}); 
for instance, we can introduce a repulsive potential for the spin up 
component of the form 
$V_\uparrow\left(z,\zeta \right)= \delta\left( z-\zeta \right)\left|\uparrow\right\rangle 
\left\langle  \uparrow \right|$ 
where $\zeta$ 
plays the role of the coordinates of a \textit{quasi--hole} in the spin up wavefunction. 
The corresponding single--particle ground--states are
\begin{equation} \label{quasiholeup}
 \phi\left( z, \zeta \right)= \mathcal{G}_1 \left(\left( z-\zeta\right) Q\left( z\right)   e^{-\frac{B}{4}\left|z\right|^2}\left|\downarrow\right\rangle\right)
\end{equation}
where $Q(z)$ is a generic polynomial. We observe that 
the wavefuntion density for the spin up component in $\phi$ 
goes to zero for $z=\zeta$ while the spin down density, in general, does not, 
due to the derivative term in $d^\dag$. 
We can also consider a repulsive potential not affected by the spin, 
$V\left( \zeta \right) =\delta\left(z-\zeta \right)$: 
in this case we obtain as a ground--state the wavefunctions
\begin{equation} \label{quasihole}
\Phi\left( z,\zeta \right) =\mathcal{G}_1 \left(\left( z-\zeta\right)^2 Q(z) e^{-\frac{B}{4}\left|z\right|^2}\left|\downarrow\right\rangle\right)
\end{equation}
having a vanishing density in $\zeta$ both for the spin up and the spin down component. 
Notice that it is impossible to create inside the space $\chi_1^-$ 
a wavefunction with a zero spin--down density 
in $\zeta$ and a non-vanishing spin up component.

With respect to the Hamiltonian (\ref{eq1}), these excitations are gapless; 
however they increase the total angular momentum of the system, and, 
as we will show in section \ref{angular}, this implies an increment in energy 
once we consider the case $\Delta = \omega- \Omega \ne 0$ in (\ref{eqH}). 

\subsection{$U(3)$ non--Abelian gauge potentials}

As shown in \cite{juze10,goldman11}, 
it is possible to engineer gauge potentials involving a
higher number of internal states. For instance, 
considering atoms with a tetrapod electronic structure, 
one can obtain three degenerate dark states, which we denote by 
$\left|+\right\rangle$, $\left|0\right\rangle$ and $\left|-\right\rangle$: 
this corresponds to an effective spin 1 and it allows 
to mimic the effect of an external $U(3)$ non--Abelian gauge potential.

In particular we can generalize the construction of the previous Section 
to the following potential:
\begin{equation} \label{su3}
A_x = -\frac{B}{2}y \,\mathbb{I}+ \begin{pmatrix}0 & \alpha & 0 \\ \alpha & 0 & \beta \\ 0 &\beta & 0 \end{pmatrix}\, , \qquad
A_y = \frac{B}{2}x \,\mathbb{I}+ \begin{pmatrix}0 & -i\alpha & 0 \\ i\alpha & 0 & -i\beta \\ 0 &i\beta & 0 \end{pmatrix}
 \end{equation}
where $\mathbb{I}$ is the $3 \times 3$ identity matrix, 
$B$ is the effective magnetic field and 
$\alpha$ and $\beta$ are two arbitrary parameters giving 
the coefficients of different Gell-Mann matrices. 
$A_x$ and $A_y$ do not commute and they 
describe a particular family of effective homogeneous non-Abelian $U(3)$ 
potentials whose corresponding non-Abelian magnetic field is
\begin{equation}
\mathcal{B}= \begin{pmatrix} B-2\alpha^2 & 0 & 0 \\ 0 & B + 2\alpha^2 - 2\beta^2 &0 \\ 0 & 0 & B+2\beta^2
             \end{pmatrix}
\end{equation}
which is translationally invariant as in Equation (\ref{eq1a}). 
These potentials are similar to the one chosen in 
\cite{goldman11} to simulate Weyl fermions through multi--component ultracold 
atoms in optical lattices.

\begin{figure}[!ht]
\centering
\includegraphics[width=7cm]{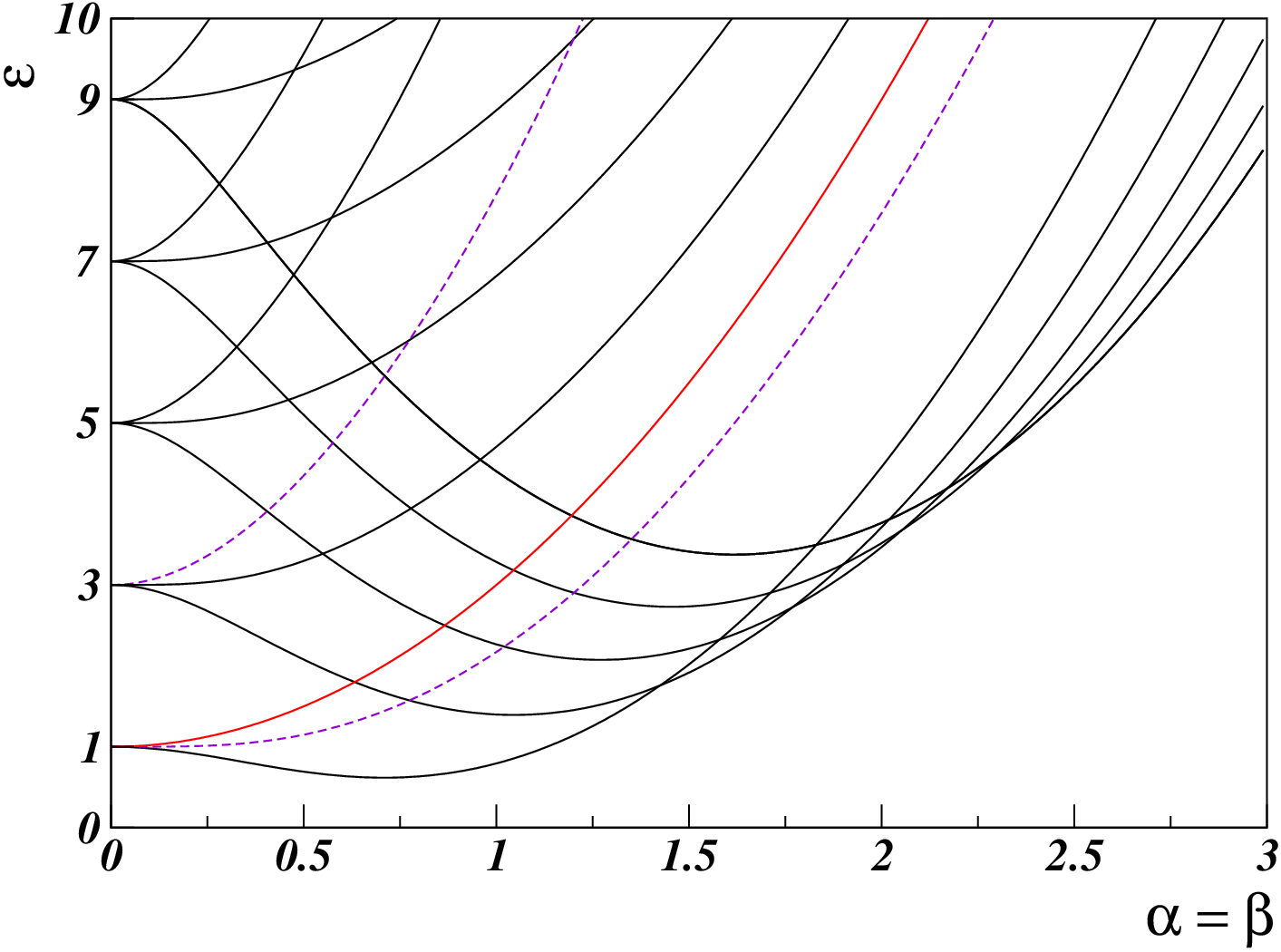}
\caption{The energy levels of the Hamiltonian (\ref{hamsu3}) 
are plotted imposing the constraint $\alpha=\beta$ (with $B=1$). 
The black lines represent the eigenvalues $\varepsilon_n$ (\ref{eigensu3}), 
the purple dashed lines represent the energy $\varepsilon_1^{\pm}$ of the
doublet states and the red line represents the energy 
$\varepsilon_0$ of the uncoupled states. 
One can see that, in the limit $\alpha=\beta \to 0$, 
the Landau level energies are recovered with a threefold pseudospin degeneracy.}
\label{figsu3}
\end{figure}

Given the potential (\ref{su3}), 
the minimal coupling Hamiltonian assumes the form:
\begin{equation} \label{hamsu3}
H\left( \alpha,\beta\right)=  B+ \frac{1}{4} d^\dag d +  
\begin{pmatrix} 2\alpha^2 & -i\alpha d & 0 \\ i\alpha d^\dag & 2\alpha^2 + 2\beta^2 & -i\beta d \\ 0 & i \beta d^\dag & 2\beta^2 \end{pmatrix}
 \end{equation}
where we used the operators $d$ and $d^\dag$ defined in (\ref{eqD}). 
The first term in $H\left( \alpha,\beta\right)$ 
is the Abelian term proportional to the identity, 
while the second one describes the non--Abelian interaction, 
depending on the parameters $\alpha$ and $\beta$, 
coupling subsequent Landau levels with different spin as in the $U(2)$ case. 
Therefore, for each $n\ge 2$, we can identify three families of eigenstates 
obtained by linear superpositions of the states 
$\psi_{n,m}\left|-\right\rangle$, 
$\psi_{n-1,m}\left|0\right\rangle$ and 
$\psi_{n-2,m}\left|+\right\rangle$, where $m$ is related to the angular momentum. 
In particular, given $n\ge 2$, 
the corresponding eigenenergies of $H\left( \alpha,\beta\right)$ 
are the solutions $\varepsilon_n$ of the following eigenvalues equation 
(see Fig. \ref{figsu3} for the case $\alpha=\beta$):
\begin{equation} \label{eigensu3}
\begin{pmatrix}
2\alpha^2 + B\left( 2n-3 \right) & -i 2\alpha \sqrt{2B\left( n-1 \right) } & 0 \\
  i 2\alpha \sqrt{2B\left( n-1 \right) } & 2\alpha^2 + 2\beta^2 + B\left( 2n-1 \right) & -i 2 \beta \sqrt{2Bn}  \\
  0 & i 2 \beta \sqrt{2Bn} & 2\beta^2 + B\left(2n+1 \right) 
 \end{pmatrix} = \varepsilon_n \mathbb{I}.
\end{equation}
Like the case of the Jaynes--Cummings coupling, 
the spin degeneracy of the Landau levels is removed and 
the eigenstates of (\ref{hamsu3}) are also eigenstates of the total angular momentum $J$.

In analogy with the previously discussed $U(2)$ potential, 
there are also other eigenstates of the Hamiltonian (\ref{hamsu3}) 
corresponding 
to the uncoupled states $\psi_{0,m}\left|-\right\rangle$ 
with energy $\varepsilon_0 = 2\beta^2 + B$ and 
to the family of (unnormalized) ``doublet states'' defined by
\begin{equation}
\Phi^{\pm}_m= \left( B- \alpha^2 \mp\sqrt{\left( \alpha^2-B\right)^2 +8B\beta^2} \right)\psi_{0,m} \left|0\right\rangle  -i\beta\sqrt{8B}\psi_{1,m}\left|-\right\rangle
\end{equation}
with energy
\begin{equation}
 \varepsilon_1^{\pm}=2B+2\beta^2+\alpha^2\pm\sqrt{\left( \alpha^2-B\right)^2 +8B\beta^2}.
\end{equation}

We observe that in the limit 
$\alpha \to 0$ or $\beta \to 0$ the results of the previous Section are 
recovered. 
In fact, if either $\alpha$ or $\beta$ goes to zero, 
the resulting gauge potential is an effective $U(2)$ potential of the kind 
(\ref{eq1}) 
and one of the spin states remains decoupled with respect to the others.

\begin{figure}[ht]
\centering
\includegraphics[width=7cm]{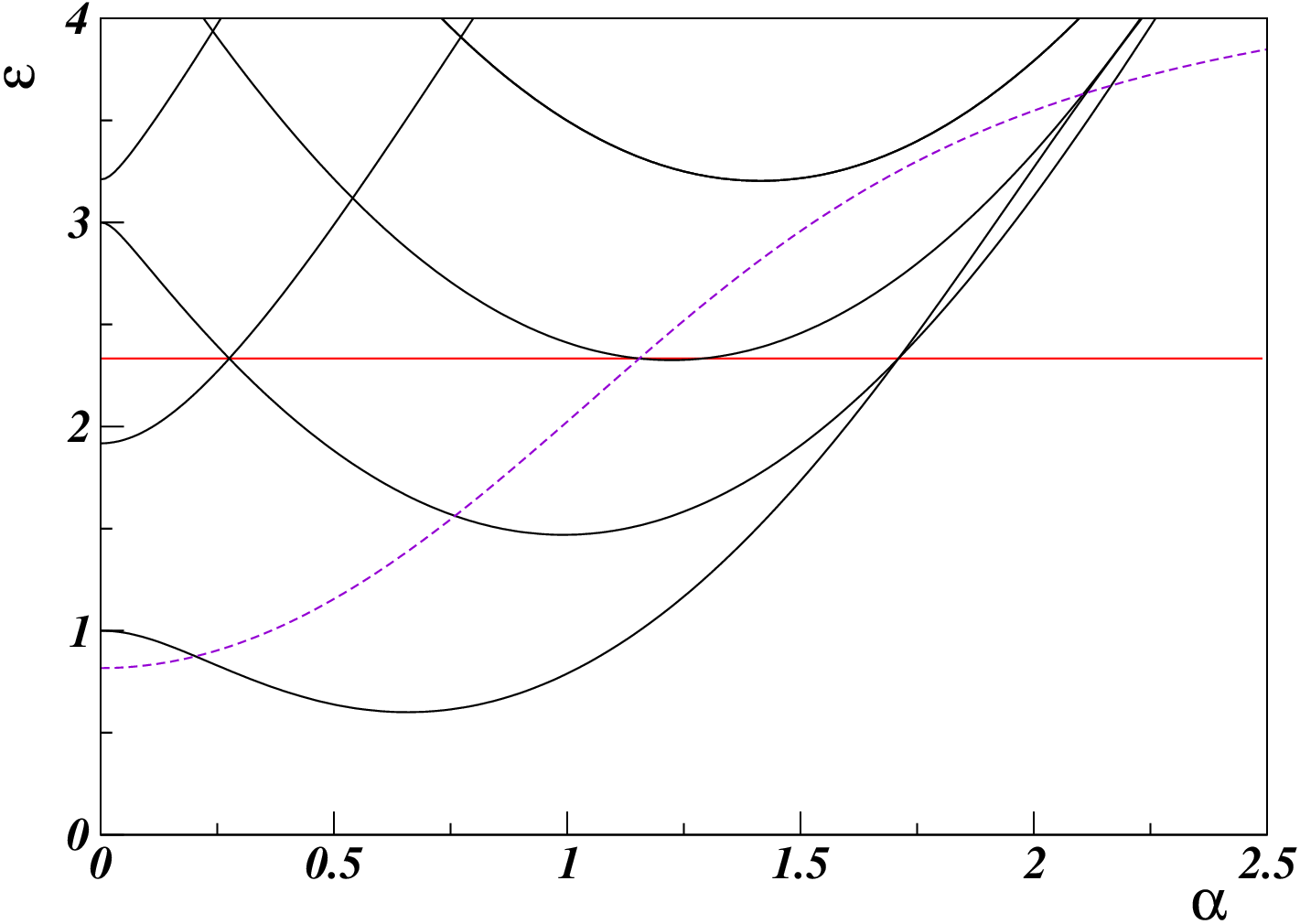}
\caption{The first energy levels of the Hamiltonian (\ref{hamsu3}) are plotted as a function of $\alpha$ for $\beta=\sqrt{2/3}$ and $B=1$. The black lines represent the eigenvalues $\varepsilon_n$ (\ref{eigensu3}), the purple dashed line represents the energy $\varepsilon_1^-$ of the first doublet states and the red line represents the energy $\varepsilon_0=7/3$ of the uncoupled states. For this choice of the parameters a triple degeneracy of the ground states appear for $\alpha=\sqrt{\left(9+\sqrt{73} \right)/6}\simeq 1.71$.
}
\label{figSU3d}
\end{figure}

Finally Fig. \ref{figSU3d} shows that it is possible to 
recover a triple degeneracy of the ground--state for particular values 
of the parameters $\alpha$ and $\beta$. 
The figure shows a triple degeneracy occurring for $B=1$, 
$\alpha=\sqrt{\left(9+\sqrt{73} \right)/6 }$ and 
$\beta=\sqrt{2/3}$ between the uncoupled state and
two eigenvectors of (\ref{eigensu3}) obtained for $n=2$ and $n=3$ 
at the energy $\varepsilon=\frac{7}{3}B$. 
We also checked that lines of doubly degenerate ground--states occur in the 
plane defined by 
$\alpha$, $\beta$ (single points with triple degeneracy belong of course to 
these lines). Further details on the spectrum 
of Hamiltonian (\ref{hamsu3}) will be presented elsewhere.

\section{Two--body interactions and deformed Laughlin states} 
\label{twobody}

In the previous Section we described a 
single particle with two internal degrees of freedom 
in the non--Abelian potential (\ref{eq1}): we consider 
in this Section a system of $N$ atoms, introducing 
two--body repulsive interactions. Denoting by $g_1$ the (dimensionless) scattering length 
between particles in the same internal state and by $g_0$ the scattering length 
between particles in different internal states, we can write the interaction 
Hamiltonian as
\begin{equation}
H_I= \sum \limits_{i<j}^N \left( g_1 \Pi_1 + g_0 \Pi_0\right) \delta \left( z_i-z_j\right).
\end{equation}
Here $\Pi_1$ is the projector over the space in which the particles $i$ and $j$ 
have parallel spin states 
($\left| \uparrow \uparrow \right\rangle$ or $\left| \downarrow 
\downarrow \right\rangle$), whereas $\Pi_0$ is the projector over the space 
in which $i$ and $j$ have antiparallel spins ($\left| \uparrow \downarrow \right\rangle$ or $\left| \downarrow \uparrow \right\rangle$). We will consider both bosonic and fermionic 
gases, keeping in mind that for fermions it is  
$g_1=0$ \cite{pitaevskii03,pethick08}.

An arbitrary two--particle state in which both atoms are in $\chi_1^-$ 
can be described as
\begin{equation} \label{eq2p}
\Psi= \mathcal{G}_{1,1} \mathcal{G}_{1,2} P\left(z_1,z_2\right) e^{-B\left(\left|z_1\right|^2+\left|z_2\right|^2 \right)/4 }\left| \downarrow \downarrow \right\rangle,
\end{equation}
where $\mathcal{G}_{1,i}$, defined in (\ref{eqG}), 
refers to the coordinate $z_i$, and $P$ is generic polynomial in $z_1$ and $z_2$. 
With vanishing inter--species interaction ($g_0=0$) and strong intra--species interaction, 
$\Psi$ has a zero interaction energy if its components 
$\left| \uparrow \uparrow \right\rangle$ , 
$\left| \downarrow \downarrow \right\rangle$ vanish when $z_1 \to z_2$: 
for fermions, this is assured by the Pauli principle; whereas for bosons  
the strong intra-species regime corresponds to $g_1 \gg B,q$ and the two--body wavefunction $\Psi$ has to fulfil the requirements 
\begin{equation}
P\left(z,z\right)=0\,,\quad \left.\left(\partial_{z_1} + \partial_{z_2}\right)P\right|_{z_1=z_2}=0\,,\quad \left.\partial_{z_1}\partial_{z_2}P\right|_{z_1=z_2}=0.
\end{equation}
Every antisymmetric polynomial $P\left(z_1,z_2\right)=-P\left(z_2,z_1\right)$ obviously satisfies these constraints, and, in general, all the fermionic functions $\Psi\left(z_1,z_2\right)$ guarantee that the intra--species interaction gives a zero energy contribution.

If we add also an inter--species repulsive interaction, such that $g_0 \gg B,q $, 
the two--particle wavefunction (\ref{eq2p}) must satisfy the further constraints
\begin{equation}
\left.\partial_{z_1}P\right|_{z_1=z_2}=\left.\partial_{z_2}P\right|_{z_1=z_2}=0
\end{equation}
in order to be a ground--state of $H_I$. These relations hold, for instance, 
in the case $P=\left(z_1-z_2\right)^m$ with $m>1$. 
In the case $m=2$ the inter--species interaction is zero, 
but not the intra--species one, 
whereas for $m\ge 3$ every repulsive potential $H_I$ gives a null contribution. 
In the following we consider the regime given by 
$g_0=0$ and (for bosons) $g_1 \gg B>3q^2$. 
Under these conditions we can generalize the previous results for the case of $N$ atoms. 

The fermionic states, antisymmetric by the exchange of every pair of atoms, 
have a zero interaction energy; thus, a possible ground--state of the 
$N$--particle Hamiltonian $\mathcal{H}=\sum_{k=1}^N H_k+H_I$, 
with all the atoms in the $\chi_1^-$ space in order 
to minimize the single--particle energy, is given by
\begin{equation} \label{eqgsa}
\Psi=\left[ {\prod\limits_i^N \mathcal{G}_{1,i}}\right] \left( \mathcal{A}\left[P_0\left(z_1 \right) ,...\,,P_{N-1}\left( z_N\right)  \right] e^{-\frac{B}{4}\sum \limits_{i} ^N \left|z_i\right|^2} \left| \downarrow \downarrow ... \downarrow \right\rangle \right), 
\end{equation}
where $\mathcal{A}$ is the antisymmetrization over all the coordinates $z_i$ and $P_m$ are different polynomials. Generalizing this kind of many-body states, it is easy to define a deformation, due to the non--Abelian potential, of the common Laughlin states. If we choose $P_m(z)=z^m$ with $m=0,...,N-1$, we obtain the usual Jastrow factor $\mathcal{A}\left[P_0...P_{N-1} \right] = \prod_{i<j}^N \left(z_i - z_j \right)$. More in general, given a Laughlin wavefunction
\begin{equation} \label{Laughlin}
\Lambda_N^{(m)}=  \prod \limits_{i<j}^N \left(z_i - z_j \right)^m  e^{-\frac{B}{4}\sum \limits_{i} ^N \left|z_i\right|^2}\left| \downarrow \downarrow ... \downarrow \right\rangle
\end{equation}
with $m$ odd, the state
\begin{equation} \label{eqgsb}
\Psi^{(m)} = \prod \limits_{j}^N \mathcal{G}_{1,j}  \Lambda_N^{(m)}
\end{equation}
is a ground--state of the Hamiltonian $\mathcal{H}$: 
every atom lies in a superposition of states $\chi_1^-$ and 
the antisymmetric wavefunction causes the intra--species interaction energy to be zero.

Also for bosons, symmetric under the exchange of two particles, there are states 
that have a zero intra-species interaction. 
For instance we can consider $\Psi^{(m)}$ for an even value of $m\ge 4$. 
In this case for each pair of particles with $z_i \to z_j$ 
the wavefunction vanishes at least as $\left(z_i-z_j\right)^2$, 
thus the interaction energy (and also its 
inter--species contribution if $g_0\neq 0$) 
is vanishing.

Therefore it is important to notice that the introduction of the 
non--Abelian gauge potential (in the regime $q^2<3B$) 
implies that the highest density deformed Laughlin state 
with null interaction energy has a filling factor $1/4$ instead of the 
usual filling factor $1/2$ that characterizes systems 
of rotating bosons \cite{cooper08,cooper05} with a contact interaction. 
Thus we expect that the introduction of the $SU(2)$ potential 
gives rise to the incompressible state $\Psi^{(4)}$, 
as numerically observed for small values of the chemical potential 
in the weak-interacting regime \cite{palmer11}. 
Such state is absent in the case of a pure magnetic field and 
it can be considered as a signature of the effect of the potential (\ref{eq1}).

The state $\Psi^{(m)}$ describes in general an incompressible fluid of 
spin--$1/2$ particles, as it can be shown calculating its norm: one finds 
\begin{multline} \label{normb}
I=\left\langle \Psi^{(m)} |\Psi^{(m)}\right\rangle =\left\langle \Lambda_N^{(m)} \right| \prod \mathcal{G}_{1,j}^\dag \mathcal{G}_{1,j} \left| \Lambda_N^{(m)}\right\rangle = \\
= \left\langle \Lambda_N^{(m)} \right| \prod \limits_j^N \left(\left|c_{\uparrow,1} \right|^2 + \left|c_{\downarrow,1} \right|^2 d_j d_j^\dag + \sigma_{x,j} \left( c_{\uparrow,1}^* c_{\downarrow,1} d_j^\dag + c_{\uparrow,1} c_{\downarrow,1}^* d_j  \right)  \right)  \left|\Lambda_N^{(m)}\right\rangle = \\
= \left(\left|c_{\uparrow,1} \right|^2 + 8B \left|c_{\downarrow,1} \right|^2\right)^N \left\langle \Lambda_N^{(m)} |\Lambda_N^{(m)}\right\rangle,
\end{multline}
where we considered that all the single--particle states involved in $\Psi^{(m)}$ 
are in the lowest Landau level and therefore are eigenstates of $dd^\dag$. 
Thus the norm $I$ can be easily written in terms of the one of the Laughlin state, 
and one can apply the 
argument also to $\Psi^{(m)}$. 

This is true also if we consider quasi--holes in the Laughlin state as, for example
\begin{equation}  \label{eq2holes}
\Psi_{\zeta_1,\zeta_2}^{(m,k)}=\prod \limits_{i}^N \mathcal{G}_{1,i} \left( \prod \limits_i^N \left( z_i -\zeta_1\right)^k \left(z_i - \zeta_2 \right)^k\right) \Lambda_N^{(m)},
\end{equation}
since each atom in the Laughlin state is in the lowest Landau level, 
and thus the operators $\mathcal{G}_1$ modify only the 
norm of the states by a constant factor for each atom.

This correspondence highlights the nature of these excitations 
since it allows us to state that the Berry phase due to the adiabatic exchange 
of the pair of quasi--holes $\zeta_1$ and $\zeta_2$ is the same of the one 
characterizing the corresponding quasi--holes in a simple Laughlin state. 
This kind of excitations are therefore Abelian anyons and their 
braiding statistics is ruled by the usual exchange properties of 
the Abelian states in the fractional quantum Hall effect \cite{arovas84}.

\subsection{Generalization to higher value of $q^2/B$} 
\label{gener}

So far we referred to the case $q^2<3B$ in which the ground--state family 
is provided by wavefunctions of the kind $\chi^-_1$. 
However, as shown in Fig. \ref{fig1}, for higher values of $q$ 
different ground--states families alternate. 
Therefore it is necessary to generalize the previous results also for $q^2 \ge 3B$ by defining the family of operators $\mathcal{G}_n$ 
describing the deformed Landau levels of the kind $\chi^-$. From (\ref{crossmm}) 
one sees that $\chi_{n>1}^-$ is the ground--state family for
\begin{equation} \label{range}
\left(2n-1 \right)B<q^2<\left(2n+1\right)B:
\end{equation}
its energy $\varepsilon_n^-$ varies, in this range of $q$, 
from $\left(2n-1 \right)B$ to $\left(2n+1\right)B$. 
To describe the ground--state family $\chi_n^-$ 
we generalize the operator $\mathcal{G}_1$ (\ref{eqG}) introducing the operators
\begin{equation} \label{Gn}
\mathcal{G}_n=c_{\uparrow,n}\,d^{\dag \, (n-1)}\sigma_x+c_{\downarrow,n}\,d^{\dag \,n},
\end{equation}
where the constants $c_{\uparrow,n}$ and $c_{\downarrow,n}$ are defined in (\ref{constc},\ref{constcb}). The ground--state wavefunctions can be expressed in the form
\begin{equation}
\chi_n^-=\mathcal{G}_n \left(P(z) e^{-\frac{B}{4}\left|z\right|^2}\left|\downarrow\right\rangle\right).
\end{equation}
Using these expressions and following the procedure shown in the case of $\chi_1^-$, 
it is possible to obtain the appropriate many--body wavefunctions for each 
value of $B$ and $q$, with $n$ chosen in order to satisfy 
(\ref{range}) (the case of the degeneracy points $q^2=\left( 2n+1\right)B$ will be analyzed in section \ref{degeneracy}). 
In particular, for an arbitrary $n$, all the antisymmetric states given by
\begin{equation} \label{generlaughlin}
 \Psi_n^{(m)} = \prod \limits_{j}^N \mathcal{G}_{n,j}  \Lambda_N^{(m)},
\end{equation}
with $\Lambda_N^{(m)}$ an odd Laughlin state (\ref{Laughlin}), 
are fermionic ground--states unaffected by repulsive intra--species contact interactions 
(here $\mathcal{G}_{n,j}$ indicates the operator $\mathcal{G}_n$ applied to the atom $j$). 

For bosons having repulsive delta interactions (both intra--species and inter--species) 
one has to consider the derivatives present in the operators $\mathcal{G}_n$ 
to find a ground--state having zero interaction energy. 
The highest order derivative in $\mathcal{G}_n$ is given by the term 
$\partial_z^n$ in the $\left|\downarrow\right\rangle$ component of each particle. 
Therefore, to identify the smallest even power $m$ in (\ref{generlaughlin}) 
annihilating a delta interaction, one has to consider 
for each pair of particles the $\left|\downarrow\downarrow\right\rangle$ 
component: 
in order to make the repulsive delta interaction null 
there must be a factor $\left(z_i-z_j\right)^m$ with $m>2n$. 
Once the polynomial order of the wavefunction (\ref{generlaughlin}) 
is high enough to make vanish the interaction in the 
$\left|\downarrow\downarrow\right\rangle$ component, 
then also all the other components give a null contribution. 
Therefore the bosonic wavefunction $\Psi_n^{(2n+2)}$ is the ground--state 
with the smallest polynomial order for generic repulsive delta interaction.

As a consequence, $\Psi_n^{(2n+2)}$ defines the maximum filling factor 
$\nu_n=1/\left(2n+2\right)$ over which the interaction energy among 
bosons cannot be zero. In the range (\ref{range}) 
only states with $\nu<1/\left(2n+2\right)$ can have a null interaction 
energy for $z_i \to z_j$ in the $\left|\downarrow\downarrow\right\rangle_{ij}$ component. 
This result is consistent with the numerical data obtained in \cite{palmer11} 
where it is shown that, in the case of $q^2<3B$, 
the Laughlin state with $\nu=1/2$ has a positive interaction energy, 
whereas for $\nu=1/4$ the energy is exactly zero. In this regime, in fact, 
$m$ must be at least $4$ to give a true ground--state, 
as already observed in the previous Section. 
Moreover, the maximum filling factor $\nu_n$ decreases as $q^2/B$ increases, therefore the role of interactions becomes more and more important 
if we consider higher filling factors for high values of $q$. 
This could explain why, for $\nu>1/2>\nu_n$, 
there are no numerical evidences of incompressible bosonic states 
after the first Landau level crossing ($q^2>3B$) \cite{palmer11}.

\section{The effect of angular momentum} 
\label{angular}

As mentioned in Section \ref{sectripod}, 
the single--particle Hamiltonian (\ref{ham}) can be considered 
as the limit of the Hamiltonian $H_L$ (\ref{eqH}) 
when the angular velocity $\Omega$ approaches the trapping frequency $\omega$. 
However, so far we considered only cases in which the condition \ref{cond2} in 
Section \ref{U2} holds and we neglected an eventual energy contribution 
of the angular momentum. In this Section we analyze the effect of the angular 
momentum term in the Hamiltonian (\ref{eqH}) 
in order to calculate the energy of the states $\Psi_n^{(m)}$ (and 
their excitations) and to determine the constraints which 
$\Delta=\omega-\Omega$ must satisfy not to spoil the Landau level description.

Let us consider the single--particle Hamiltonian (\ref{eqH}):
\[
H_L=\left( \vec p+\vec A\right) ^2 -  \Delta  L_z.  
\]
The term proportional to $\Delta$ is spin--independent and it does not 
affect the non-Abelian contribution $H_{na}$ in (\ref{eqhna}). 
Using its eigenstates $\varphi_{n,m}^\pm$ (\ref{eqc}) for $n\ge 1$ 
we can split $H_L$ into blocks of the form
\[
H_{L,n,m}=2q^2+2Bn-\Delta\left( n-m-\frac{1}{2}\right) +  \begin{pmatrix}   2q \sqrt{2Bn}   & -B +\frac{\Delta}{2} \\ -B +\frac{\Delta}{2} &  - 2q \sqrt{2Bn} 
\end{pmatrix}.
\]
The eigenenergies of $H_L$ are therefore
\begin{equation} \label{energyl}
\varepsilon_{n,m}^{\pm}=2Bn+2q^2- \Delta\left( n-m-\frac{1}{2}\right)  \pm\sqrt{\left(B-\frac{\Delta}{2}\right)^2+8q^2Bn } 
\end{equation}
and the corresponding (unnormalized) eigenstates are 
\begin{multline} \label{chidelta}
 \chi_{n,m}^\pm=\left( B - \frac{\Delta}{2}+ 2q \sqrt{2Bn} \mp \sqrt{\left(B-\frac{\Delta}{2}\right)^2+8q^2Bn}\right) \psi_{n-1,m}\left|\uparrow\right\rangle +\\
+ \left( B - \frac{\Delta}{2} - 2q \sqrt{2Bn } \pm \sqrt{\left(B-\frac{\Delta}{2}\right)^2+8q^2Bn}\right) \psi_{n,m}\left|\downarrow\right\rangle.
\end{multline}
We see that the uncoupled eigenstates are unaffected 
by the angular momentum term. 

Since the coefficients in the definition of the eigenstates do not depend on $m$, 
then one can redefine the constants $c_{\uparrow,n}$, $c_{\downarrow,n}$ 
and the operators $\mathcal{G}_n$ independently on $m$. Moreover, 
the Landau level structure holds also for $H_L$ (condition \ref{cond1}), 
and the Landau levels are energetically distinguishable 
provided that the energy contribution of the angular momentum 
remains small with respect to the Landau level spacing. 
In fact, all the energy levels (\ref{energyl}) have a term 
which is linear in the total angular momentum $J=n-m-1/2$ 
(increasing their energy if $J<0$). Thus the states with 
lower values of $m$ are favoured: this implies an energy gap $\Delta$ 
for the creation of single--particle quasi--holes of the kind (\ref{quasiholeup}) 
and $2\Delta$ for (\ref{quasihole}) (for $\zeta=0$).

In order to understand the stability of the Landau level structure, 
and thus of the deformed Laughlin states, we have to consider a 
multiparticle wavefunction describing $N$ atoms in a ground--state of the 
kind $\Psi_n^{(m)}$ (\ref{generlaughlin}) 
for $2n-1<q^2/B<2n+1$. In this case a particle corresponding 
to the highest value of the modulus of the angular momentum term $\left| J\right| $ 
acquires an additional energy $\sim Nm\Delta$ that must be smaller 
than the gap with the next deformed Landau level. 
This gap, for values of $q^2$ far from the degeneracy points, 
can be calculated evaluating the energy difference of $\chi^-_n$ 
and $\chi^-_{n-1}$ in the crossing point between 
$\chi^-_{n-1}$ and $\chi^-_{n+1}$ and it can be approximated by $B/(2n)$. 
Therefore the Landau level description of the multiparticle states 
remains accurate if $\Delta \ll B/(2nNm)$.

Let us analyze more in detail the regime characterized by small values of $\Delta$. 
The deformed Laughlin states $\Psi_n^{(m)}$ have to be defined 
with the appropriate corrections in the operators 
$\mathcal{G}_n$ (\ref{Gn}) since the constants 
$c_\uparrow$ and $c_\downarrow$ must include $\Delta$ coherently with (\ref{chidelta}). 
The corresponding energy contribution of the total angular momentum turns to be 
for $\Psi_n^{(m)}$
\begin{equation}
-\Delta J \Psi_n^{(m)} =  \Delta N\left( \frac{mN}{2}-m -n +\frac{1}{2} \right)   \Psi_n^{(m)} \approx \Delta m \frac{N^2}{2} \Psi_n^{(m)}.
\end{equation}
Therefore, the Laughlin states with smaller $m$ (and higher density) 
are energetically favoured: thus, in the case of fermions, 
far from the degeneracy points, the ground--states are described 
by the $\nu=1$ filling factor states $\Psi_n^1$, 
whereas in the case of bosons the ground--states are of the form $\Psi_n^{2n+2}$.

A quasi--hole in the position $\zeta=0$
\begin{equation}
\Psi_n^{(m)} = \prod \limits_{j}^N \mathcal{G}_{n,j} \prod \limits_{j}^N z_j^k \Lambda_N^{(m)}
\end{equation}
acquires an additional energy $N\Delta k$ with respect to the corresponding 
ground--state, because it changes the angular momentum of each particle by $k$. 
This energy can be considered the gap for the creation of a quasi--hole.

\section{Degeneracy points and non--Abelian anyons} 
\label{degeneracy}

An important property of the single--particle spectrum of the Hamiltonian 
(\ref{ham}) is the existence of degeneracy points 
corresponding to the values $q^2=\left(1+2n \right)B$: this makes possible 
the occurrence of many--particle ground--states having non--Abelian excitations 
\cite{trombettoni10}. In these degeneracy points 
the two lowest energy levels cross 
(possibly generating a first-order phase transition \cite{palmer11}) 
and the ground--state degeneracy of the single--particle is doubled. 
In these points an atom in the (non--interacting) ground--state 
can be described by all the superpositions of wavefunctions in 
$\chi_n^-$ and $\chi_{n+1}^-$. However, if we consider the angular 
momentum term in the Hamiltonian (see the previous Section), 
then the states with a lower angular energy are favoured and the variation of the parameter $q$ around the degeneracy points gives rise to a crossover, as we will describe in \ref{sec:crossover}.

When the intra--species interaction between atoms is introduced, 
the doubled degeneracy of the single--particle states 
implies a novel form of the multiparticle ground--state 
which is quite different from (\ref{eqgsa},\ref{eqgsb}). 
We first consider the case of the first degeneracy point, $q^2=3B$, 
in which there is the crossing between a ground--state 
of particles in $\chi_1^-$ and $\chi_2^-$ 
(our conclusions will be later extended to all the other degeneracy points). 
For the single particle the basis of ground--states is defined by the set 
$\left\lbrace \mathcal{G}_1 z^m \psi_0\left( z\right) , \mathcal{G}_2 z^m 
\psi_0\left( z\right),\; {\rm with} \; m\in \mathbb{N}\right\rbrace $ and,  analogously to the previous sections,
we have to distinguish the case of interacting bosons 
and the one of (free) fermions.

\subsection{Fermionic gases}

We first analyze the fermionic case: the highest density 
ground--state function of the Hamiltonian $\mathcal{H}$ 
for $2N$ atoms is given by \cite{trombettoni10}
\begin{equation} \label{eqgscr}
\Omega_c= \mathcal{A}\left[\mathcal{G}_1 \psi_0, \mathcal{G}_1 z \psi_0,...\,, \mathcal{G}_1 z^{N-1} \psi_0,\, \mathcal{G}_2 \psi_0, \mathcal{G}_2 z \psi_0,...\,, \mathcal{G}_2 z^{N-1} \psi_0     \right]   \left| \downarrow \downarrow ... \downarrow \right\rangle, 
\end{equation}
where $\mathcal{A}$ implements the full antisymmetrization over all the atoms. 
Because of the double degeneracy, 
the wavefunction $\Omega_c$ describes an atomic gas with 
filling factor $\nu=2$. 
The state $\Omega_c$ is obtained through the Slater determinant 
of the single--particle wavefunctions with the lowest angular momenta 
$\left|J\right|$, up to the power $z^{N-1}$. Therefore, 
considering also the angular momentum contribution in the Hamiltonian, 
it is the true ground--state for the system.

The double degeneracy makes $\Omega_c$ very different from the case 
(\ref{eqgsa}): in the degeneracy points 
the antisymmetrization hides a clustering of the particles 
into two sets of $N$ atoms, say $A$ and $B$, 
that are physically different and refer to states in $\chi_1^-$ and in 
$\chi_2^-$. The two clusters must have the same number of atoms 
in order to minimize the contribution to the energy given 
by the angular momentum. The previous wavefunction can 
be recast in the form
\begin{equation} \label{eqcluster}
\Omega_c=\mathcal{A}\left[ \prod \limits_{k \in A} \mathcal{G}_{1,k} \prod \limits_{i<j \in A} \left( z_i - z_j\right)  \prod \limits_{l \in B} \mathcal{G}_{2,l} \prod \limits_{i<j \in B} \left( z_i - z_j\right)      \right] e^{-\frac{B}{4}\sum \limits_{i} ^{2N} \left|z_i\right|^2} \left| \downarrow \downarrow ... \downarrow \right\rangle
\end{equation}
where we made explicit the Jastrow factor 
in each cluster and $\mathcal{A}$ refers to the antisymmetrization over 
all the possible clusterings in the sets $A$ and $B$. 
In the spirit of the quantum Hall states 
showing a clustering into two sets - 
the main example being the Moore and Read (MR) Pfaffian state 
\cite{moore91,nayak96} - 
$\Omega_c$ is characterized by the presence of 
quasi--hole excitations corresponding 
to half a quantum flux (and effective charge $1$, since $\nu=2$). 
These excitations must appear in pairs: 
it is however helpful to analyze the two possible wavefunctions 
they can assume. Such wavefunctions are related to 
quasi--holes in the two different clusters, given by 
\begin{eqnarray} \label{eqholes2a} 
\sigma_{1}\left(\zeta \right) =\mathcal{A}\left[ \prod\limits_{A} \mathcal{G}_{1,k}   \prod \limits_{A}  \left( z_i - z_j\right)  \prod\limits_{A}\left(z_i-\zeta \right) \prod\limits_{B} \mathcal{G}_{2,l} \prod \limits_{B} \left( z_i - z_j\right) \right] e^{-\frac{B}{4}\sum \limits_{i} ^{2N} \left|z_i\right|^2} \left| \downarrow ... \downarrow \right\rangle \\
\sigma_{2}\left(\zeta \right) =\mathcal{A}\left[ \prod\limits_{A} \mathcal{G}_{1,k}   \prod \limits_{A}  \left( z_i - z_j\right)   \prod\limits_{B} \mathcal{G}_{2,l} \prod \limits_{B} \left( z_i - z_j\right) \prod\limits_{B}\left(z_i-\zeta \right) \right] e^{-\frac{B}{4}\sum \limits_{i} ^{2N} \left|z_i\right|^2} \left| \downarrow ... \downarrow \right\rangle \label{eqholes2b}.
\end{eqnarray}
These quasi--holes obey a fermionic statistics: once two of them 
of the same kind are exchanged, the wavefunction 
acquires a $\pi$ phase. However, it is interesting to notice 
that they show the same fusion rules of the Ising model 
with defined fermionic parity \cite{georgiev09} 
(characterizing the MR state \cite{moore91,nayak96}) 
once one defines a third bosonic excitation given 
by the fusion $\psi \equiv \sigma_1 \times \sigma_2$. 
From the physical point of view, {\it if} it is possible to obtain 
linear superpositions of $\sigma_1$ and $\sigma_2$ 
through the interplay between repulsive potentials for the 
$\left|\uparrow\right\rangle $ and $\left|\downarrow\right\rangle $ 
components, then the so--obtained (non--Abelian) 
excitations could present interesting features, a point which 
certainly deserves further investigations.


In fermionic systems at the degeneracy point $q^2=3B$, 
the ground--state $\Omega_c$, characterized by the filling factor $\nu = 2$, 
is the highest density state obtained with atoms in the Hilbert space 
spanned by the deformed Landau level $\chi_1^-$ and $\chi_2^-$. 
Moreover, $\Omega_c$ minimizes the term in the Hamiltonian proportional 
to the angular momentum. Nevertheless, 
in the study of rotating ultracold atomic gases, 
it is interesting to analyze what happens varying the filling factor, 
since, in general, such systems present 
non-trivial phase diagrams as a function of $\nu$ 
\cite{cooper01,cooper05,cooper08}. 
As we have already shown, the double degeneracy 
at this particular value of $q^2$ provides in a natural way a
 clustering of the atoms into two sets in order to minimize 
$\left|J \right|$. Each atom can assume a wavefunction 
which is a superposition of states in $\chi_1^-$ and in $\chi_2^-$ and 
transitions from one to the other are possible: 
therefore, also for smaller values of the filling factor, 
we are driven to consider deformed ground--states 
showing a pairing among tha atoms that are similar 
to the ones usually considered in the study of 
fractional quantum Hall effect \cite{readgreen}.

Let us consider first the filling factor $\nu=1$: 
in this case a paired state is built 
by favouring the creation of coupled atoms 
in the antisymmetric state obtained 
by applying the operator 
$\left(\mathcal{G}_{1,i}\mathcal{G}_{2,j}-
\mathcal{G}_{2,i}\mathcal{G}_{1,j}\right)$ to the pair of atoms 
$(i,j)$ in the limit $z_i \to z_j$. 
The corresponding wavefunction for $2N$ particles reads
\begin{equation} \label{omega} 
\Omega_{\rm Hf} = {\rm Hf}\left( \left( \mathcal{G}_{1,i}\mathcal{G}_{2,j}-G_{2,i}G_{1,j}\right) \frac{1}{z_i-z_j}\right)\prod\limits_{i<j}^{2N}\left( z_i-z_j\right) e^{-\frac{B}{4}\sum \limits_{i} ^{2N} \left|z_i\right|^2} \left| \downarrow \downarrow ... \downarrow \right\rangle   
\end{equation}
where $\rm Hf$ indicates the Haffnian, which is a symmetric version of the Pfaffian defined for a symmetric matrix $M_{ij}=M_{ji}$:
\begin{equation} \label{sympfaf}
 {\rm Hf}\left( M\right)=\sum\limits_{\sigma \in \mathcal{P}} M_{\sigma(1),\sigma(2)} M_{\sigma(3),\sigma(4)}... M_{\sigma(2N-1),\sigma(2N)} 
\end{equation}
(the sum is over all the permutation of the indices). 

The wavefunction $\Omega_{\rm Hf}$ is 
antisymmetric over all the atoms because it is composed 
by the symmetric Haffnian and by the antisymmetric Jastrow factor. 
This implies that intra--species interactions give a zero 
contribution to the energy and $\Omega_{\rm Hf}$ can be considered a 
ground--state since each atom lies 
in a superposition of states of $\chi_1^-$ and $\chi_{2}^-$.  
However this wavefunction is not vanishing for $z_i \to z_j$ 
because of the components with different spin, 
therefore $\Omega_{\rm Hf}$ is not in general a 
ground--state for inter--species repulsive interactions that 
require higher powers of the Jastrow factor to give a null contribution.


For $\nu = 1/2$ there are two possible antisymmetric paired 
states that have been widely analyzed in the literature. 
The first one corresponds to a deformed MR Pfaffian state 
\cite{moore91,nayak96} and the second corresponds to a deformed Haldane-Rezayi 
state \cite{haldane,moore91}. 
The deformed MR Pfaffian state can be described 
by an effective $p$-wave pairing \cite{readgreen} 
obtained by applying the operator 
$\left(\mathcal{G}_{1,i}\mathcal{G}_{2,j}+
\mathcal{G}_{2,i}\mathcal{G}_{1,j}\right)$ 
which favours a symmetric state 
(with respect to $\chi_1^-$ and $\chi_{2}^-$) 
for the pair $(i,j)$ when $z_i \to z_j$. The corresponding wavefunction is 
\cite{trombettoni10}
\begin{equation} \label{mr} 
\Omega_{MR} = {\rm Pf}\left( \left( \mathcal{G}_{1,i}\mathcal{G}_{2,j}+\mathcal{G}_{2,i}\mathcal{G}_{1,j}\right) \frac{1}{z_i-z_j}\right)\prod\limits_{i<j}^{2N}\left( z_i-z_j\right)^2 e^{-\frac{B}{4}\sum \limits_{i} ^{2N} \left|z_i\right|^2} \left| \downarrow \downarrow ... \downarrow \right\rangle   
\end{equation}
where $\rm Pf$ is the Pfaffian operator. 
This wavefunction is antisymmetric, 
therefore intra--species interactions give a null contribution and 
$\Omega_{MR}$ can be considered a ground--state since each 
atom lies in a superposition of states in $\chi_1^-$ and $\chi_{2}^-$. 
This state shares all the main characteristics of the Moore and Read 
wavefunctions \cite{moore91}, and, in particular, 
its excitations are non--Abelian Ising anyons, 
as shown in \cite{nayak96} where an analogous wavefunction is analyzed. 
$\Omega_{MR}$ can be mapped into the usual spinless MR state $\Psi_{MR}$ 
\cite{moore91,nayak96} in a way which is similar to equation 
(\ref{eqgsb}) for the ground--state outside the degeneracy points. 
Since, for every factor in the Pfaffian, one atom is in a state 
in $\chi^-_1$ and the other in $\chi^-_2$, the norm of 
$\Omega_{MR}$ is obtained from the one of the Pfaffian state $\Psi_{MR}$ 
just by multiplying it by a constant value for each pair of atoms:
\begin{equation} \label{eqmr2}
\left\langle \Omega_{MR}\right|\left. \Omega_{MR}\right\rangle =\left(\left|c_{\uparrow,1}\right|^2+8B\left|c_{\downarrow,1}\right|^2 \right)^N \left( 8B\left|c_{\uparrow,2}\right|^2+2(8B)^2\left|c_{\downarrow,2}\right|^2\right) ^N\left\langle \Psi_{MR}\right|\left. \Psi_{MR}\right\rangle.
\end{equation}
This constant value can be calculated in a way similar to equation 
(\ref{normb}) and it is an effect of the clustering characterizing 
the state $\Omega_{MR}$. Equation (\ref{eqmr2}) guarantees 
that also the statistics of the excitations $\zeta_a$ and $\zeta_b$ 
of the kind
\begin{equation} 
\Omega_{MR}\left(\zeta_a,\zeta_b \right) = {\rm Pf}\left( \frac{\mathcal{G}_{1,i} \mathcal{G}_{2,i} \left(z_i-\zeta_a \right) \left(z_j-\zeta_b \right) + i\leftrightarrow j }{z_i-z_j}\right)\prod\limits_{i<j}^{2N}\left( z_i-z_j\right)^2 e^{-\frac{B}{4}\sum \limits_{i} ^{2N} \left|z_i\right|^2} \left| \downarrow \downarrow ... \downarrow \right\rangle   
\end{equation}
can be described by Ising anyons as in the case of the Moore and Read state, 
since the Berry and monodromy phases acquired in 
the exchange of two excitations coincide.

The other paired ground--state at $\nu=1/2$ is a deformed Haldane-Rezayi 
state that can be obtained through the introduction 
of the antisymmetric operator 
$\left(\mathcal{G}_{1,i}\mathcal{G}_{2,j}-\mathcal{G}_{2,i}\mathcal{G}_{1,j}\right)$:
\begin{equation} \label{hr} 
\Omega_{HR} = {\rm Pf}\left( \left( \mathcal{G}_{1,i}\mathcal{G}_{2,j}-\mathcal{G}_{2,i}\mathcal{G}_{1,j}\right) \frac{1}{\left( z_i-z_j\right)^2}\right)\prod\limits_{i<j}^{2N}\left( z_i-z_j\right)^2 e^{-\frac{B}{4}\sum \limits_{i} ^{2N} \left|z_i\right|^2} \left| \downarrow \downarrow ... \downarrow \right\rangle.   
\end{equation}
This state has a total angular momentum $\left|J\right|$ 
which is lower than the one of $\Omega_{MR}$ so that, in principle, 
it is energetically favoured if we consider the single-particle 
Hamiltonian (\ref{eqH}). However the Haldane-Rezayi state 
represents the critical point between a weak and a strong coupling 
phase in a fermionic system with an effective $d$-wave pairing 
\cite{readgreen}, therefore it is considered to be a gapless state 
whose excitations are described by a non--unitary conformal field 
theory \cite{gurarie} which is unfit to define an incompressible state. Moreover we expect that $\Omega_{HR}$ is more influenced by the presence of a weak inter--species interaction than $\Omega_{MR}$.

To conclude the discussion about the (free) fermionic gases 
it is worth noticing that all the states presented can 
be retrieved also for the generic degeneracy point 
between states in $\chi^-_n$ and $\chi^-_{n+1}$ 
by substituting the operators $\mathcal{G}_1$ and $\mathcal{G}_2$ 
with $\mathcal{G}_n$ and $\mathcal{G}_{n+1}$. This is true in the 
case of free fermions (i.e., with $g_0=0$), whereas the 
introduction of strong interactions, 
such as an inter--species contact repulsion, 
brings to different scenarios presenting deformed Halperin states, 
similar to the one we will present for the bosonic gases, 
whose exponents and filling factors depend also on the value of $n$.

\subsection{Bosonic gases}

The analysis of the degenerate points can be applied also to bosonic gases. 
As described in Section \ref{gener}, 
the state $\Psi^{(m)}_n$ (\ref{generlaughlin}) 
is a ground--state of both the intra--species and the inter--species contact interactions for every $m>2n$, 
therefore $\Psi^{(2n+2)}_n$ is the bosonic wavefunction 
with zero interaction energy minimizing the polynomial order. 
In the point $q^2=3B$ one has a superposition of states in 
$\chi_1^-$ and in $\chi_2^-$ and, as in the previous case, 
we can describe the related multiparticle 
ground--state through the clustering in two corresponding subsets $A$ and $B$. 
The resulting ground--state is
\begin{multline} \label{bosonic}
\Omega_b=\mathcal{S}\left[ \prod \limits_{k \in A} \mathcal{G}_{1,k} \prod \limits_{i<j \in A} \left( z_i - z_j\right)^4  \prod \limits_{l \in B} \mathcal{G}_{2,l} \prod \limits_{i<j \in B} \left( z_i - z_j\right)^6 \prod \limits_{k\in A, l\in B} \left(z_k-z_l \right)^4 \right] e^{-\frac{B}{4}\sum \limits_{i} ^{2N} \left|z_i\right|^2} \left| \downarrow \downarrow ... \downarrow \right\rangle  = \\
= \mathcal{S} \left[ \prod \limits_{k \in A} \mathcal{G}_{1,k} \prod \limits_{l \in B} \mathcal{G}_{2,l} \prod \limits_{i<j \in B} \left( z_i - z_j\right)^2\right] \Lambda^{(4)}_{2N} 
\end{multline}
where the symmetrization over all the atoms is necessary 
to have a bosonic state and $\Lambda^{(4)}$ is the generalized 
Laughlin state (\ref{Laughlin}). 
The exponents of the Jastrow factors are defined 
considering that the lowest polynomial order term 
for an atom in $A$ is given by the first derivative 
included in $\mathcal{G}_1$, whereas in $B$ 
by the second derivative in $\mathcal{G}_2$. 
Therefore $\Omega_b$ vanishes at least as 
$\left( z_i -z_j\right)$ whenever $z_i \to z_j$ 
for each pair of atoms, and the intra--species interaction energy is zero. 
$\Omega_b$ is characterized by a filling factor $\nu = 2/9$ and 
it is built by the symmetrization of the Halperin state 
$\Psi\left(4,6,4\right)$. The interaction term between atoms 
in different clusters is determined in order to satisfy the 
zero interaction energy constraint
\[
\prod \limits_{k\in A, l\in B} \partial_{z_k} \partial^2_{z_l} \left(z_k-z_l \right)^4 \propto \left(z_k-z_l \right).
\]
The $\Omega_b$ filling factor, $\nu=2/9$, is therefore the highest possible filling factor that guarantees a null interaction energy for bosons at the point $q^2=3B$. 

We conclude this Section observing that considering different 
degeneracy points, $q^2=(2n+1)B$, the ground states for intra--species 
repulsions are defined as deformed Halperin states 
$\Psi\left(2n+2,2n+4,2n+2\right)$ characterized by lower filling factors.

\subsection{Crossover at a degeneracy point} \label{sec:crossover}

So far we considered the non--Abelian component of the potential 
at the exact value $q^2 = (2n+1)B$ which is characterized 
by a perfect degeneracy of the deformed Landau levels 
$\chi_n^-$ and $\chi_{n+1}^-$. Then the population of atoms 
must be equally distributed into these levels 
in order to minimize the total angular momentum of the system and 
thus its total energy. However, if the parameter $q^2$ 
is slightly detuned from these degeneracy points, 
the deformed Landau level with a lower energy 
will present an increase in population 
balancing the small energy difference arising between the two state families.

Let us consider, in particular, a system showing a 
filling factor $\nu_n = 1/m_n$ on the $n^{\rm th}$ Landau level, 
and let us suppose that $q^2$ is slightly higher than $(2n+1)B$, 
so that $\varepsilon_{n,m}^- > \varepsilon_{n+1,m}^-$. 
To fill the gap between $\chi_n^-$ and $\chi_{n+1}^-$ 
there must be an imbalance $M=N_{n+1}-N_{n}$ 
between the population of atoms in $\chi_{n+1}^-$ and $\chi_{n}^-$, such that 
\[
\varepsilon_{n+1,\,M m_{n+1}}^- \approx \varepsilon_{n,0}^-.
\] 
Considering the energy eigenvalues given by Equation (\ref{energyl}), 
one obtains
\begin{equation}\label{emme}
M m_{n+1} \Delta \approx \delta q^2\, \left.\frac{d\left( \varepsilon_{n,0}^- - \varepsilon_{n+1,0}^- \right)}{dq^2} \right|_{q^2=(2n+1)B} \approx \left( \frac{1}{4n+3} + \frac{1}{4n+1}\right) \delta q^2,
\end{equation}
where $\delta q^2$ is the displacement of $q^2$ 
from the degeneracy point: we assumed $\Delta \ll B$ 
in the second approximation done in Equation (\ref{emme}). 
In order to obtain a system described by the deformed Hall states 
defined above for the degeneracy point at $q^2=3B$, 
the imbalance $M$ must be negligible 
with respect to the total number of atoms $N$ and, 
in particular, one obtains the following degeneracy 
condition on the displacement $\delta q^2$ around $3B$:
\begin{equation} \label{crossover}
 \frac{12}{35}\left| \delta q^2 \right| \ll \frac{N}{\nu} \Delta < \left(9 - \sqrt{73} \right)B \approx 0.46 B, 
\end{equation}
where $\nu$ is the total filling factor of the system and 
the further constraint for $N\Delta$ is derived from the 
energy difference with the third Landau level, $\chi_3^-$, 
in such a way that all the atoms lie only in the two lower Landau levels. 
If $\delta q^2$ exceeds the limit (\ref{crossover}), 
the imbalance $M$ grows up until only the population 
in the lower Landau level is left. 
Therefore the displacement $\delta q^2$ 
drives a crossover between different regimes 
characterized by wavefunctions in $\chi_n^-$ and $\chi_{n+1}^-$ 
such as the deformed Laughlin states in (\ref{generlaughlin}) 
and only in the regime defined by the condition (\ref{crossover}) 
Hall states presenting particles in both 
the deformed Landau levels can be present.

\section{Conclusions}
\label{conclusions}

We studied two--component ultracold bosonic and fermionic atomic gases in 
$U(2)$ non--Abelian potentials. We focused our attention 
to gauge potentials having, in the Abelian limit, doubly 
degenerate Landau levels deriving from an Abelian magnetic field equal for 
both the components. We investigated the effect of general 
homogeneous non--Abelian terms, 
discussing the conditions under which the structure of degenerate 
Landau levels is preserved (even if the spin degeneracy is split 
by the coupling between the internal states realized by the non--Abelian term). 
We argued that there are only three classes of (quadratic) 
Hamiltonians preserving the degeneracy of the Landau levels and 
giving rise to an analytically defined Landau level structure, 
where the eigenstates are expressed as a finite linear combination 
of the eigenstates of a particle in a magnetic field. 
The first class corresponds to an Abelian $U(1)\times U(1)$ 
gauge potential and it refers to uncoupled internal states, 
whereas the other two are characterized by a truly non--Abelian 
$U(2)$ gauge potential and correspond to different kinds 
of Jaynes--Cummings models. 

Focusing on one of these Jaynes--Cummings classes which 
is gauge equivalent both to a Rashba and to a Dresselhaus 
spin--orbit coupling, we determined the single--particle energy spectrum 
and we discussed the parameters of the Rabi pulses needed for the 
physical implementation of such gauge potentials. It is found 
that the single--particle energy levels assume the same values 
in a series of degeneracy points. 
The corresponding deformed quantum Hall states for fermions and bosons 
(in the presence of strong intra--species interaction) 
have been determined through a mapping to the usual quantum Hall Laughlin 
states. 
Such mapping preserves the Berry phases characterizing the 
exchanges of quasi--holes and, therefore, the Abelian excitation statistics. 
Far from the degeneracy points, deformed Laughlin states
are found; whereas, at the crossing points of the lowest Landau levels, 
ground--states 
with non--Abelian excitations emerge at different fillings. 
A detailed discussion of the properties 
of the resulting deformed Moore--Read states and of the state $\Omega_c$ 
(minimizing the total angular momentum) was presented; besides, 
the crossover arising at the degeneracy points was investigated, estimating 
the range of the Hamiltonian parameters for the validity of these 
many--body ground states. 

The effect of an angular moment term on the stability of the Landau level 
structure and of the deformed Laughlin states was analyzed and we provided 
an estimate of the values of the angular momentum such that the Landau level 
description of the multiparticle states remains accurate. 
We also gave results for $U(3)$ gauge potentials 
acting on a three--component gas, pointing out that it is possible 
to have lines of degeneracy points of the Landau levels 
and triple degeneracies in the spectrum.

It would be interesting in the future to study the transport 
properties of the quantum Hall states found for ultracold atoms in $U(2)$ 
artificial gauge potentials preserving the Landau levels, especially in view 
of the experimental study of the signatures of the different Hall states. 
From this point of view, we mention that it could be useful 
to study setups in which the Abelian terms are 
different for the two components. Another related 
issue, important for the experimental detection 
of non--Abelian states, would be also the investigation of 
the physical addressability of such states, e.g. 
how the filling of these states 
depends on the experimental parameters. Equally important would be the estimation of Haldane 
pseudopotentials relevant to address the effect of more general interacting terms like inter--species repulsions and dipolar long--range interactions. 

\textit{Acknoledgements} 
We thank N. Barber\'an, B. Estienne, G. Mussardo, J. Pachos, R. Palmer, K. Schoutens and X. Wan for very useful discussions and correspondence. 
This work has been supported by the grants INSTANS (from ESF) and 2007JHLPEZ (from MIUR).


\begin{thebibliography}{XX}
\bibitem{note1}
The difference in sign with respect to \cite{ruse05} is related to the 
different convention chosen for the Hamiltonian.
\end{thebibliography}
\end{document}